\documentclass[reprint,onecolumn,eqsecnum,amsmath,nofootinbib]{revtex4-2}

\usepackage{graphicx}
\usepackage{bm}
\usepackage{booktabs}

\newcommand\apjs{Astrophysical Journal Supplement Series}
\newcommand\mnras{Monthly Notices of the Royal Astronomical Society}

\usepackage[colorlinks=true, allcolors=blue]{hyperref}

\begin{document}

\title{A post-Newtonian approach to neutron star oscillations}

\author{Shanshan Yin}
\affiliation{Mathematical Sciences and STAG Research Centre, University of Southampton, Southampton SO17 1BJ, United Kingdom}
\author{Nils Andersson}
\affiliation{Mathematical Sciences and STAG Research Centre, University of Southampton, Southampton SO17 1BJ, United Kingdom}
\author{Fabian Gittins}
\affiliation{Institute for Gravitational and Subatomic Physics (GRASP), Utrecht University, Princetonplein 1, 3584 CC Utrecht, Netherlands}
\affiliation{Nikhef, Science Park 105, 1098 XG Amsterdam, Netherlands}

\begin{abstract}

Next-generation gravitational-wave detectors are expected to constrain the properties of extreme density matter via observations of static and dynamical tides in binary neutron star inspirals. The required modelling is straightforward in Newtonian gravity---where the tide can be represented in terms of a sum involving the star's oscillation modes---but not yet fully developed in general relativity---where the mode-sum approach is problematic. As a step towards more realistic models, we are motivated to explore the post-Newtonian (pN) approach to the problem (noting that the modes should still provide an adequate basis for a tidal expansion up to 2pN order). Specifically, in this paper we develop the pN framework for neutron star oscillations and explore to what extent the results remain robust for stars in the strong-field regime. Our numerical results show that the model is accurate for low-mass stars ($\lesssim 0.8M_{\odot}$), but becomes problematic for more massive stars. However, we demonstrate that the main issues can be resolved (at the cost of abandoning the consistency of the pN expansion) allowing us to extend the calculation into the neutron star regime. For canonical neutron stars ($\approx 1.4M_\odot$) our adjusted formulation provides the fundamental mode of the star with an accuracy comparable to that of the relativistic Cowling approximation. For lower mass stars, our approach is significantly better.

\end{abstract}

\maketitle

\section{Introduction}

The properties of cold dense matter above the nuclear saturation density remain relatively poorly constrained by experiment and astrophysical observations (see \cite{2024LRR....27....3K} for a recent review). As a result, the precise nature of the matter deep inside neutron stars is still uncertain. This uncertainty is typically encoded in the equilibrium equation of state for matter, the relation providing the thermodynamic pressure as a function of energy density (temperature, matter composition, etcetera). This microscopic relation is required to determine the macroscopic properties---essentially, the mass-radius relation---of neutron stars (via the Tolman-Oppenheimer-Volkoff equations) which can then be tested against observational data.  The current state-of-the-art for tests based on electromagnetic observations draws on data from  NASA’s Neutron Star Interior Composition Explorer (NICER) mission \cite{2019Miller,2019Riley, 2024Salmi, 2024Dittmann}.

Gravitational-wave observations of compact binaries involving at least one neutron star open another promising avenue for exploration. Specifically, observational constraints on the tidal deformability \cite{2008Flanagan,Chatziioannou2020}---encoding the response of a neutron star to the tidal interaction induced by a binary companion---provide information on the neutron star radius. The tightest such constraints to date were obtained for GW170817 \cite{2017Abbott-GW170817,2018Abbott.PhysRevLett.121.161101}, the first observed binary neutron star system, with weaker constraints inferred from the subsequent GW190425  event \cite{2020Abbott}. The current error bars on the  neutron star radius inferred from gravitational-wave data are similar to those gleaned from NICER. 

Future gravitational-wave observations, with more sensitive instruments like the Einstein Telescope \cite{2020Maggiore} and Cosmic Explorer \cite{2019BAAS...51g..35R}, are expected to lead to significantly tighter constraints (see, for instance, the discussion in \cite{2022PhRvD.105h4021C}). In addition to probing the mass-radius relation, we expect to gain insight into the composition and state of matter in the neutron star core. This involves important issues like the presence of hyperons and/or deconfined quarks, various macroscopic superfluid/superconducting components and so on. In order to explore these aspects, we need to look behind the static tidal deformability and consider the impact of dynamical tides \cite{ 2016PhRvL.116r1101H, 2016PhRvD..94j4028S, 2020Andersson, 2020NatCo..11.2553P}.
During a binary inspiral, the tidal field of the companion induces a time-varying mass-quadrupole moment, deforming the star, enhancing the gravitational-wave emission and accelerating the coalescence. 
As the orbital separation approaches the size of the bodies, the details of the stars’ internal structure become important. In particular, the tidal force excites individual stellar oscillation modes---as they become resonant with the tidal driving \cite{1994LAi,1995MNRAS.275..301K}--- leading to additional transfer of orbital energy into the stellar fluid, potentially leaving a observable impact on the orbital evolution. This is the dynamical tide. 

The dynamical tide manifests in a number of ways. The dominant contribution is associated with the star's fundamental f-mode \cite{ 2016PhRvL.116r1101H, 2016PhRvD..94j4028S, 2020Andersson}. This mode may not reach actual resonance before merger, but its presence nevertheless leads to a notable enhancement of the tidal response. Models that do not include this enhancement introduce an unnecessary systematic error in the extracted neutron star parameters (and hence the equation of state constraints) \cite{2020NatCo..11.2553P}. At a more subtle level, a number of low-frequency modes may be dynamically excited as the system sweeps through the sensitivity band of a detector. The most likely such resonances are associated with gravity
modes linked to the matter composition  \cite{2023PhRvD.108d3003H} and possible interface modes associated with internal phase transitions (e.g. the crust-core transition \cite{2012PhRvL.108a1102T,2020PhRvL.125t1102P,2023PhRvD.107h3023Z} or a first order phase transition to an exotic matter phase at higher density) \cite{2024ApJ...964...31M,interface_paper}. The problem is even richer for rotating stars. Not only does stellar rotation shift the mode resonances (e.g. making the f-mode resonance more likely before merger) \cite{2021PhRvR...3c3129S,2024MNRAS.527.8409P,2024PhRvD.110b4039Y}, it also brings new modes into existence. The most promising of these inertial modes is thought to be the r-modes \cite{2000Lockitch_PhysRevD.63.024019, 2023Andersson_rmode}, which couples to the tide gravitomagnetically \cite{2007PhRvD..75d4001F,2020PhRvD.101j4028P}. 

The possibility that dynamical tide features may be within reach of observations motivates a detailed analysis for next-generation detectors, like the Einstein Telescope and Cosmic Explorer. In parallel with the design of these instruments, we need to improve our theoretical models. This involves adding all relevant aspects (or, at least, as many as we can manage...) of neutron star physics.  %\cite{2021Poisson.PhysRevD.103.064023,2021Kuan10.1093/mnras/stab1898}. 
A key part of this effort involves improving on the Newtonian mode-sum approach, which is commonly used to model dynamical tides \cite{1994LAi,1995MNRAS.275..301K}. The underlying idea is simple.
If the oscillation modes of a star form a complete set then they can be used as a basis to represent the tidal deformation as a sum of modes. In Newtonian gravity this is the case \cite{1978ApJ...221..937F,2001PhRvD..65b4001S}. However, it is not expected to remain true in full general relativistic calculations \cite{1967Thorne,1969Price.Thorne-II,1969Thorne-III,1969Thorne-IV,1983Lindblom,1985Detweiler}. The main difficulty in general relativity is that the modes are not going to be complete (due to the existence of late-time power-law tails associated with wave scattering by the curved spacetime) and the gravitational-wave damping also makes the problem non-Hermitian \cite{2021Andersson}. This presents a technical challenge if we want to use realistic matter physics in our models for the dynamical tide \cite{2024PhRvD.109f4004P, 2024Hegade.PhysRevD.109.104064}.

A possibly way to progress the discussion would be to explore the problem within post-Newtonian theory. 
While post-Newtonian (pN in the following) models are not expected to be very accurate for relativistic stars one would expect them to be decidedly better than Newtonian ones. In particular, one may build post-Newtonian models for realistic equations of state \cite{2023Andersson}. In addition, one should be able to develop a post-Newtonian mode-sum strategy for tides as the set of stellar oscillation modes is believed to remain complete up to at 2pN order (see the arguments in \cite{gittins2025perturbationtheorypostnewtonianneutron}). It would seem, at least conceptually, relevant to pursue this strategy. Ultimately, the effort may only represent a small step towards a more accurate description of dynamical neutron-star tides but we still hope to learn useful lessons from the analysis.  
As a step in this direction, we calculate neutron star oscillation modes in post-Newtonian theory in this paper. 

The paper is organised as follows. To begin with, in section~\ref{Sec_background} we build  neutron stars in post-Newtonian theory. We then formulate the post-Newtonian oscillation problem in section \ref{Sec_pNoscillation} and discuss our numerical results in section~\ref{Sec_results}. Finally, we conclude in section \ref{Sec_summary}. Unless otherwise indicated we will be using geometric units in which $c=G=1$.

\section{Building post-Newtonian neutron stars} \label{Sec_background}

The definitive treatise of the modern approach to the post-Newtonian method was provided by Poisson and Will about a decade ago \cite{2014grav.book.....P}. They describe the foundations of the approach and its links to both Newtonian gravity and Einstein's theory. This exhaustive text provides a natural starting point for a range of relevant applications. In particular, the problem of tides raised in compact binaries is laid out in detail. Having said that, there is still room for the development of applications connecting with other aspects of physics. For the tidal problem, a natural question relates to modern matter equations of state obtained from nuclear physics arguments. This is the issue that motivates  the effort we present here. We want to explore to what extent we can make progress on modelling dynamical neutron star tides within the pN framework. The reason for exploring this question is not an expectation that a pN model would be exceptionally precise. Not at all! We are quite realistic in this respect. However, it is well known that the problem of fully relativistic dynamical tides involves a number technical challenges (see \cite{2024PhRvD.109f4004P} for a recent discussion). Intuitively, one might expect some of these issues to be absent in (low-order) pN theory. For example, the formulation of a mode-sum for the tidal response---the go-to approach in Newtonian tidal theory that is not expected to be (at least not easily) extendable to general relativity---can be formulated in the pN framework \cite{gittins2025perturbationtheorypostnewtonianneutron}. Given this, it is interesting to pursue the problem and see how far we get.

An important feature of a pN model is that, by including matter terms of order $1/c^2$, we can allow  for the internal energy and hence work with realistic neutron star equations of state. There is, however (and famously!), no such thing as a free lunch. The main problem we face is hard wired into the pN strategy. Notably, in post-Newtonian hydrodynamics \cite{2014grav.book.....P} it is customary to carry out the calculation in such a way that equations are truncated at a specific order in a $1/c^2$ expansion, yet allowing some higher order terms to remain. As long as we operate in the strict weak-field regime, the presence of these higher order terms is irrelevant. They are all small. However, neutron stars have moderate to strong internal gravitational fields so the higher terms that are kept in the calculation will impact on the results. This is (obviously) not a nice feature. Having said that, the  alternative---to carry out a strict order by order pN expansion (as in the calculations reported in \cite{1991NYASA.631...97C})---is not an attractive proposition either (for reasons explained in \cite{gittins2025perturbationtheorypostnewtonianneutron}). At the end of the day, we need to work through the calculation if we want to assess the quality of the results. In particular, building on the work in \cite{2023Andersson} on static post-Newtonian fluid configurations and our recent analysis of the Hermitian properties of pN fluid perturbations in \cite{gittins2025perturbationtheorypostnewtonianneutron} we thus set out to formulate and solve the problem of calculating oscillation modes in the pN framework.  The calculation we carry out shares many aspect with the recent work in \cite{2023Boston}. The one important distinction is that, while that work applies the method to white dwarfs (where the pN approach should be ``safe''), our intention is to push the calculation into the strong-field regime relevant for neutron stars to find out if, when and how it breaks.

In order to solve the perturbation problem, we need to start from a suitable background configuration. 
Our previous work \cite{2023Andersson} provides useful models in this respect, but also raises the warning flag we have already alluded to. Because of the inclusion of higher order pN terms, there is a significant degree of freedom in building stellar models. The presence of higher order terms dictates (quite naturally) the density at which a given model ``breaks down'', but  some formulations perform better than others. This introduces a certain element of ``black magic'' (not an uncommon feature in approximation theory) which we need to keep in mind in the following. 

Having made these cautionary remarks, we consider two of the models formulated in \cite{2023Andersson}. Our first model, from now on referred to as PW, builds on the discussion in \cite{2014grav.book.....P}. The second model, which we will refer to as AGYM, involves a physically motivated reformulation found to lead to  more accurate neutron star models in \cite{2023Andersson}. 

The background stars are spherically symmetric, static  configurations involving a perfect fluid described by the mass density $\rho$, the internal energy density per unit mass $\Pi$ and the pressure $p$.
In the PW model the gravitational potential, $U$, is sourced by the baryon mass, $M_B$, which satisfies
\begin{equation} \label{PW_MB}
    \cfrac{dM_B}{d r} = 4\pi \rho^* r^2  \ ,
\end{equation}
 which leads to
\begin{equation} \label{PW_U}
    \cfrac{dU}{d r} = - \cfrac{GM_B}{r^2} \ .
\end{equation}
Moreover,  $\rho^*$ is a rescaled mass density \cite{2014grav.book.....P},  defined by
\begin{equation}
    \rho^* = \rho\left(1+\cfrac{3U}{c^2}\right) \ .
\end{equation}
It is also worth noting that---as is commonly the case in post-Newtonian models---the radial distance $r$ is expressed in the isotropic coordinates \cite{2023Andersson}. We also need the mass $\mathcal{N}$, contributing at 1pN order, given by
\begin{equation} \label{PW_N}
    \cfrac{d \mathcal{N}}{d r} = 4\pi \rho^* r^2  \left(  \Pi - U  + \cfrac{3p}{\rho^*} \right) \  ,
\end{equation}
where the fluid's internal energy per unit mass $\Pi$ is related to the total energy density $\varepsilon$ through
\begin{equation} \label{introduce_internal_energy}
    \varepsilon = \rho c^2  + \rho \Pi \ .
\end{equation}
With these definitions, the equation for hydrostatic equilibrium takes the form
\begin{equation} \label{PWpressure}
    \cfrac{d p}{d r} = - \cfrac{G \rho^*}{r^2} \left\{ M_B +   \cfrac{1}{c^2} \left[ \left( \Pi - 3U  + \cfrac{p}{\rho^*}  \right)M_B + \mathcal N \right] \right\} \ .
\end{equation}
It is worth noting that that, as $\rho^*$ is involved in the equations, the calculation is not consistently truncated at  1pN order.

To build a pN neutron star model, we pick an initial value for the potential, $U(r=0)=U_0$, at the centre, integrate the above equations to the point where the pressure vanishes $p(R) = 0$. This defines the stellar surface, $r=R$. Then we execute a root search to determine $U_0$ by matching the potential at the surface to the exterior in such a way that
\begin{equation} \label{BC}
    U(R) = \cfrac{GM_B(R)}{R} \ .
\end{equation}

In contrast, the AGYM model combines variables in such a way that the gravitational potential is sourced by the gravitational mass $M$ (as would be the case in relativity). To effect this, we introduce  the mass $M$ as
\begin{equation}
   M = M_B + \cfrac{1}{c^2}\mathcal N \ , 
\end{equation}
which then satisfies
\begin{equation}
    \cfrac{d M}{dr} = 4\pi \rho^* \left[1 + \cfrac{1}{c^2} \left( \Pi - U  + \cfrac{3p}{\rho} \right)  \right] r^2  \ . 
\end{equation}
and leads to the gravitational potential being determined by 
\begin{equation}
    \cfrac{dU}{dr} = - \cfrac{GM}{r^2} \ .
\end{equation}
By discarding some (not all!) 2pN terms from \eqref{PWpressure}, we can write the hydrostatic equilibrium equation in the alternative form
\begin{equation} \label{Alter2pressure}
    \cfrac{d p}{d r} = - \cfrac{G \rho^*}{r^2} \left[ M +   \cfrac{1}{c^2} \left( \Pi - 3U  + \cfrac{p}{\rho^*}  \right)M_B \right]   \approx - \cfrac{GM \rho}{r^2} \left[ 1 + \cfrac{1}{c^2} \left( \Pi + \cfrac{p}{\rho} \right)\right] =  - \cfrac{GM}{r^2 c^2}  (p+\varepsilon) \ .
\end{equation}
For the AGYM model the surface boundary condition changes to 
\begin{equation}
    U(R) = \cfrac{GM(R)}{R} \ .
\end{equation}
Finally, if we want to compare to relativistic models obtained from the standard Tolman-Oppenheimer-Volkoff equations then 
the radius in isotropic coordinates can be easily transformed to  Schwarzschild coordinates  through
\begin{equation} 
    R_S = R \left(1+\frac{GM}{2Rc^2}\right)^2 \ .
\end{equation}

\begin{figure}
    \centering
    \includegraphics[width=0.75\textwidth]{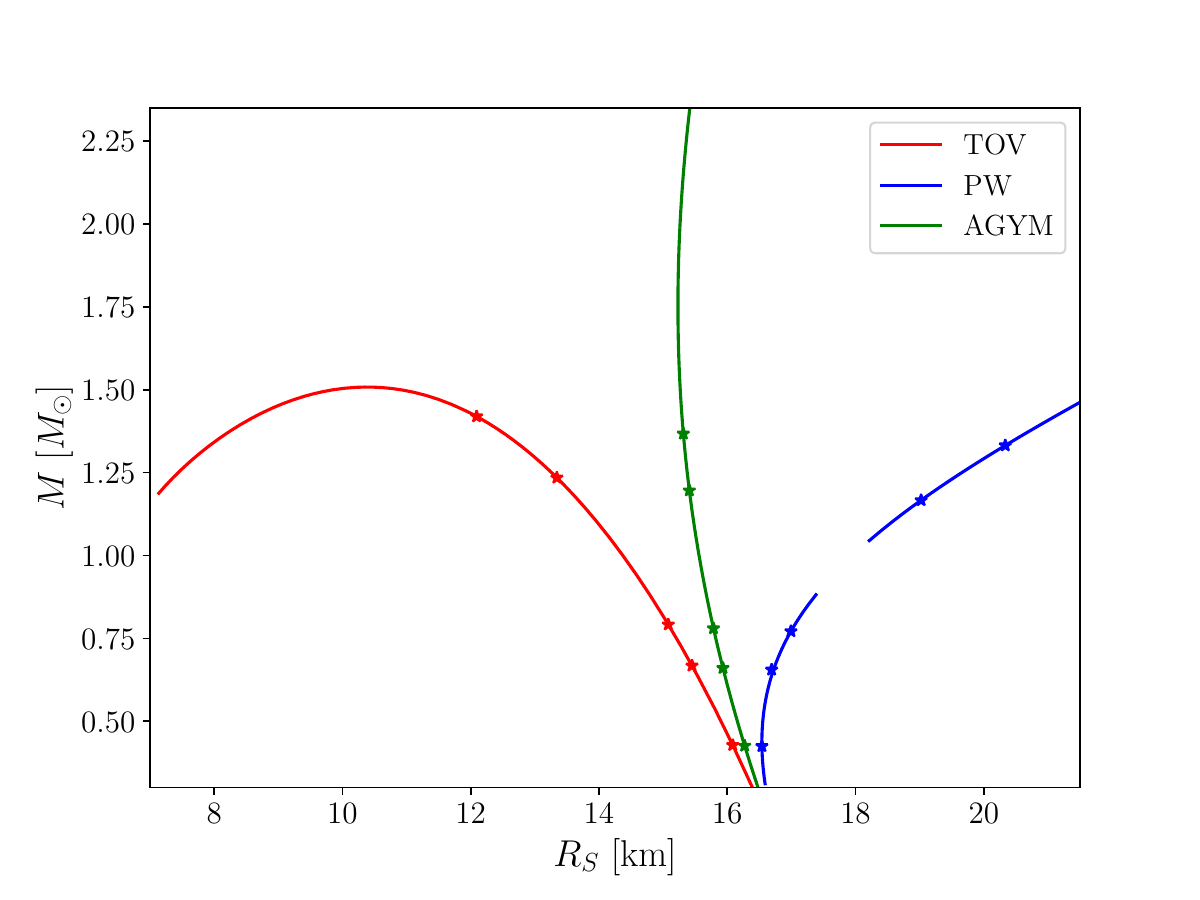} 
    \caption{Mass-radius curves for polytropic post-Newtonian stars determined from \eqref{PWpressure} (blue solid) and \eqref{Alter2pressure} (green solid). For comparison we also show  the results for relativistic stars built by solving the Tolman-Oppenheimer-Volkoff equations (red). The gap appearing in the PW curve is the region where we are not able to build any stars (for more details, see the discussion in \cite{2023Andersson}). Specific pN models considered in the mode calculations later are indicated on each respective mass-radius curve. }
    \label{PWfig}
\end{figure}

The two formulations, PW and AGYM, notably differ only in terms that contribute beyond 1pN order. Nevertheless, we know from the discussion in \cite{2023Andersson} that they lead to rather different mass-radius relations for neutron star densities. These results were established for an equation of state based on nuclear physics arguments (specifically from the BSk family of models \cite{2013Fantina, 2013Goriely.PhRvC..88b4308G}). As the present analysis is more at the level of a proof-of-principle, here we instead focus on a  phenomenological polytropic matter model. That is, we use 
\begin{equation} \label{polytrope_eos}
    p = K \rho^{\Gamma} \ ,
\end{equation}
where $\rho$ is the rest mass density, $K$ is a constant and $\Gamma=1+1/n$ with $n$ the polytropic index. Combined with the thermodynamical relation \eqref{thermrel} the internal energy $\Pi$ then has the form 
\begin{equation}
    \Pi = \cfrac{K \rho^{\Gamma -1}}{\Gamma - 1}  \ .
\end{equation}
The specific stellar models we consider follow from using $K=185$~km$^2$ and $\Gamma=2$, which allows for the existence of a solution to the Tolman-Oppenheimer-Volkoff equations with the canonical neutron star mass of $1.4M_\odot$, see Figure~\ref{PWfig}. However, the parameters are  not particularly realistic because the maximum mass reached by the model is far too low. This lack of realism is not a major concern for us, though, because the results in Figure~\ref{PWfig} also show that the pN configurations deviate from the relativistic results already at lower densities. Our main interest is in the phenomenology. We want to see if the ``problematic'' features of the background models are inherited by the perturbations and  explore to what extent additional complications enter the oscillation problem. 

\section{The non-radial oscillation problem} \label{Sec_pNoscillation}

Having obtained suitable background models, we want to study non-radial perturbations and calculate the star's oscillation modes. Given the issues associated with the background configuration, this may not be entirely straightforward. With this in mind, we will pay attention to the details of each required step.

\subsection{The perturbation equations} \label{Subsec_perturb_eq}

The derivation of the pN perturbation equations is, in principle, straightforward. As the background star is static, it is natural to work with Eulerian perturbations (the change of a physical quantity at a fixed point in space) of the various thermodynamical quantities ($\delta p$, $\delta \rho$, and so on) alongside the Lagrangian displacement vector $\xi^i$, in our case simply given by
\begin{equation}
    \delta v^i = \partial_t \xi^i \ .
\end{equation}

Starting from the momentum equation for pN hydrodynamics, it is easy to show that the perturbations must satisfy (see \cite{1965Chandrasekhar,2014grav.book.....P,gittins2025perturbationtheorypostnewtonianneutron})
\begin{multline}
    \rho^* \left\{ \left[1 + \cfrac{1}{c^2} \left(  \Pi + 3 U  + \cfrac{p}{\rho^*}  \right) \right]  \partial_t^2 \xi ^i   - \cfrac{4}{c^2} \partial_t^2 \delta V^i - \cfrac{1}{2 c^2} \partial_t^2 \partial^i \delta X  \right\} \\
    +  \left( 1 + \cfrac{2U}{c^2} \right)  \left( \partial^i \delta p - \cfrac{\delta \rho^*}{\rho^*} \partial^i p  \right)
    - \cfrac{1}{c^2}  \left(\delta \Pi  + \cfrac{\delta p}{\rho^*} - \cfrac{p}{\rho^{*2}} \delta \rho^* \right) \partial^i p  \\
    -  \cfrac{1}{c^2} \left( \partial^i p - {4\rho^*} \partial^i U \right) \delta U 
    - \rho^* \left[1 + \cfrac{1}{c^2} \left( \Pi - U  + \cfrac{p}{\rho^* }  \right) \right]  \partial^i \delta U    - \cfrac{1}{c^2} \rho^* \partial^i \delta \psi
    = 0  
\end{multline}
along with the perturbed continuity equation
\begin{equation}
    \delta \rho^* = -\partial_i(\rho^*\xi^i) \ .
\end{equation}
The perturbations of the various potentials are governed by
\begin{equation} \label{nabla_delta_U}
    \nabla^2 \delta U = -4\pi G\delta\rho^* \ ,
\end{equation}
\begin{equation} \label{perturbed_delta_Ui}
    \nabla^2 \delta V_i = -4 \pi G\rho^*\xi_i \ ,
\end{equation}
\begin{equation}
    \nabla^2 \delta\psi = -4\pi G\rho^*\left(-\delta U+\delta\Pi+ \cfrac{3 \delta p}{\rho^*} - 3p \cfrac{\delta \rho^*}{\rho^{*2}}\right)  -4\pi G\delta\rho^*\left(-U+\Pi+\cfrac{3p}{\rho^*} \right) \ ,
\end{equation}
and
\begin{equation} \label{nabla_delta_X}
    \nabla^2 \delta X = 2\delta U \ .
\end{equation}

Finally, from the pN gauge condition  (see  \cite{2014grav.book.....P})
\begin{equation}
    \partial_t U + \partial_j U^j = 0 \ , 
\end{equation}
we also have
\begin{equation}
    \partial_t \delta U + \partial_j \delta U^j = 0 \ ,
\end{equation} 
with $\delta U^j = \partial_t \delta V^j$.
This completes the set of equations we need to solve. 

\subsection{Separation of variables}

In order to solve the perturbation equations, we first of all work in the frequency domain, i.e. assume that all perturbations depend on time as $e^{i\omega t}$, so that we have
\begin{equation}
    \delta v^i = i\omega \xi^i
\end{equation}
Secondly, we note that the equations we wrote down are in Cartesian coordinates (using partial derivatives). This is, however, not a natural choice for stars; we clearly want to work in spherical coordinates and make use of angular harmonics to decouple the equations. There are two ways to effect the required change. We can either proceed in a fully covariant way, replacing partial derivatives with covariant ones and so on, or we can take a short cut. Opting for the latter (as it provides a more immediate route to the answer and also allows for a direct comparison with most of the work on the Newtonian oscillation problem, see \cite{1979Unno}) we simply reinstate the basis vectors and write the equations as (with vectors indicated as bold)
\begin{multline}
    -\omega^2 \rho^* \left\{ \left[1 + \cfrac{1}{c^2} \left(  \Pi + 3U  + \cfrac{p}{\rho^*}  \right) \right]  \bm{\xi}  - \cfrac{4}{c^2} \delta  {\bm{V} }- \cfrac{1}{2 c^2} \bm{\nabla} \delta X  \right\} \\
    +  \left( 1 + \cfrac{2U}{c^2} \right)  \left( \bm{\nabla} \delta p - \cfrac{\delta \rho^*}{\rho^*} \bm{\nabla} p  \right)  
    - \cfrac{1}{c^2}  \left(\delta \Pi    + \cfrac{\delta p}{\rho^*} - \cfrac{p}{\rho^{*2}} \delta \rho^* \right) \bm{\nabla} p \\
    -  \cfrac{1}{c^2} \left( \bm{\nabla} p - {4\rho^*} \bm{\nabla} U \right) \delta U  
    - \rho^* \left[1 + \cfrac{1}{c^2} \left(  \Pi - U  + \cfrac{p}{\rho^*}  \right) \right]  \bm{\nabla} \delta U    - \cfrac{1}{c^2} \rho^* \bm{\nabla} \delta \psi
    = 0   \ , 
\end{multline}
with 
\begin{equation}
    \delta \rho^* = - \bm{\nabla} \cdot (\rho^* \bm{\xi} ) 
\end{equation}
and 
\begin{equation}
\nabla^2 \delta  {\bm{V}} = -4  \pi G \rho^*  \bm{\xi} \ .
\end{equation}
For these equations, we have everything we need from the corresponding Newtonian problem (see, for example, the discussion in \cite{1979Unno}), apart from the form for the Laplacian of the vector $ \delta \bm{V}$. The missing information is, however, readily available from, for example, the viscous term in the text-book version of the Navier-Stokes equations in spherical coordinates. With this in hand, we are ready to proceed. 

The next step involves expanding all scalar variables ($\delta p, \ \delta U, \ \delta \psi, \ \delta X$) in spherical harmonics $Y^m_l$ in such a way that 
\begin{equation} 
    \delta p(t,r,\theta, \phi) = e^{i\omega t} \sum_l \delta p_l(r) Y^m_l(\theta, \phi)  \ , 
\end{equation}
and similarly for the other variables. Meanwhile, we use that standard decomposition for the displacement vector $\bm{\xi}$, i.e. in an orthonormal spherical coordinate basis $(\hat{\bm{r}}, \hat{\bm{\theta}}, \hat{\bm{\phi}})$, we have
\begin{equation} 
    \bm{\xi} = \xi^r \hat{\bm{r}} + \xi^{\theta} \hat{\bm{\theta}} + \xi^{\phi} \hat{\bm{\phi}}  = e^{i\omega t} \sum_l \left[ \xi_l^r(r), \xi_{l}^h(r)\cfrac{\partial}{\partial \theta}, \xi_l^h(r) \cfrac{\partial}{\sin\theta \partial\phi} \right] Y_{l}^m(\theta, \phi) \ ,
\end{equation}
where $\xi^r_l$ and $\xi^h_l$ denote the radial and tangential components, respectively. The expansion for the perturbed vector potential $\delta \bm{V}$  takes the same form (with components $\delta V_l^r$ and $\delta V_l^h$).

Noting that we need a matter equation of state to close the systems of equations, we introduce an adiabatic index $\Gamma_1$ associated with the perturbations. This assumes that the oscillations take place adiabatically and that relevant nuclear reactions are sufficiently slow that the matter composition may be considered frozen (see \cite{2024MNRAS.531.1721C} for a recent discussion)
\begin{equation} \label{adidatic}
    \cfrac{\Delta p}{p} = \Gamma_1 \cfrac{\Delta \rho}{\rho} \ ,
\end{equation}
where $\Delta$ represents a Lagrangian perturbation, which associates the small change of a quantity to a specific fluid element. This relation allows us to relate $\delta \rho_l$ to $\delta p_l$ via
\begin{equation}
    \delta \rho_l = \cfrac{\rho}{\Gamma_1} \cfrac{\delta p_l}{p} - \rho \xi_l^r A \ ,
\end{equation}
where the Schwarzschild discriminant $A$ takes the usual form
\begin{equation}
    A = \cfrac{d \ln \rho}{d r} - \cfrac{1}{\Gamma_1} \cfrac{d \ln p}{d r} \ . 
\end{equation}
Therefore, $\delta \rho_l^*$ can be eliminated in favour of $\delta p$ using:
\begin{equation}
    \delta \rho_l^* = \delta\rho_l \left(1+\cfrac{3U}{c^2}\right) + \cfrac{3\rho}{c^2}\delta U_l 
    = \cfrac{\rho^*}{\Gamma_1} \cfrac{\delta p_l}{p} - \rho^* \xi_l^r A   + \cfrac{3\rho}{c^2}\delta U_l  \ .
\end{equation}

For the internal energy---for both barotropic and non-barotropic perturbations, see Appendix~\ref{thermal_perturbation}---we have 
\begin{equation} \label{deltaPi_internal_energy}
    \delta \Pi_l = \cfrac{p}{\rho^2} \delta \rho_l = \cfrac{p}{\rho^2} \left( \cfrac{\rho}{\Gamma_1} \cfrac{\delta p_l}{p} - \rho \xi_l^r A \right) = \cfrac{\delta p_l}{\Gamma_1\rho} -  \cfrac{p}{\rho}\xi_l^r A \ .
\end{equation}

We proceed to split the equation of  motion in radial and tangential components, using the equation of the state to eliminate $\delta \rho_l, \ \delta \rho_l^*$ and $ \delta \Pi_l$. Thus,  we arrive at a set of ordinary differential equations (for each $l$-multipole)
\begin{multline}  \label{pN_eq_perturbed_Euler_radial}
    -\omega^2 \rho^*  \left[1 + \cfrac{1}{c^2} \left( \Pi + 3 U  + \cfrac{p}{\rho^* }  \right) \right] \xi^r_l + \left( 1+ \cfrac{2U}{c^2} \right) A \cfrac{d p}{d r}\xi^r_l \\
     + \left( 1+ \cfrac{2U}{c^2} \right)   \cfrac{d \delta p_l}{d r} - \left( 1 + \cfrac{2U}{c^2} \right)   \cfrac{1}{\Gamma_1 p} \cfrac{d p}{d r}\delta p_l - \cfrac{1}{c^2 \rho^*}  \cfrac{d p}{d r} \delta p_l   \\
    -  \cfrac{4}{c^2} \left( \cfrac{d p}{d r} - \rho^* \cfrac{d U}{d r}  \right) \delta U_l  - \rho^* \left[1 + \cfrac{1}{c^2} \left(  \Pi - U  + \cfrac{p}{\rho^* }  \right) \right]  \cfrac{d \delta U_l}{d r}  \\
    + \omega^2  \cfrac{4\rho^*}{c^2} \delta V^r_l - \omega^2   \cfrac{\rho^*}{2 c^2} \cfrac{d}{dr} \delta X_l   - \cfrac{1}{c^2} \rho^* \cfrac{d}{dr} \delta \psi_l    = 0  \ ;
\end{multline}
and
\begin{multline}  \label{pN_eq_Perturbed_Euler_perp}
    \omega^2 \left[1 + \cfrac{1}{c^2} \left(  \Pi + 3U  + \cfrac{p}{\rho^* }  \right) \right] \left\{  - \rho^* A  \xi^r_l  +  \cfrac{d\rho^*}{d r} \xi^r_l + \cfrac{\rho^* }{r^2} \cfrac{d}{d r} (r^2 \xi^r_l) \right\}  \\
    + \omega^2 \left[1 + \cfrac{1}{c^2} \left(  \Pi + 3U  + \cfrac{p}{\rho^* }  \right) \right]  \cfrac{\rho^*}{p \Gamma_1} \delta p_l  - \left( 1+\cfrac{2U}{c^2} \right) \cfrac{l(l+1)}{r^2} \delta p_l \\
     - \omega^2 \cfrac{\rho^*}{c^2} \delta U_l + \rho^* \left[1 + \cfrac{1}{c^2} \left(  \Pi - U  + \cfrac{p}{\rho^*}  \right) \right]  \cfrac{l(l+1)}{r^2} \delta U_l 
     - \omega^2  \cfrac{4\rho^*}{c^2} \cfrac{1}{r^2}\cfrac{d}{d r}\left(r^2\delta V^r_l \right) \\
     + \omega^2 \cfrac{\rho^*}{2 c^2} \cfrac{l(l+1)}{r^2} \delta X_l + \cfrac{\rho^*}{c^2}  \cfrac{l(l+1)}{r^2} \delta \psi_l  = 0 \ . 
\end{multline}

Similarly, the perturbation equations of the potentials, obtained from \eqref{nabla_delta_U}-\eqref{nabla_delta_X}, can be written 
\begin{equation} \label{pN_eq_delta_U}
    \cfrac{1}{r^2} \cfrac{d}{d r} \left( r^2 \cfrac{d \delta U_l}{d r} \right) -\cfrac{l(l+1)}{r^2} \delta U_l + 4\pi G \delta \rho^*_l = 0  \ ,
\end{equation}
\begin{equation} \label{pN_eq_delta_hat_Ur}
    \cfrac{1}{r^2} \cfrac{d}{d r}\left(r^2 \cfrac{{d} \delta V^r_l}{{d} r}\right) + \cfrac{2}{r} \cfrac{d \delta V^r_l}{d r}  + \cfrac{2 - l(l+1)}{r^2}  \delta V^r_l   + \cfrac{2}{r} \delta U_l    + 4 \pi G\rho^* \xi^r_l = 0  \ ,
\end{equation}
\begin{multline} \label{pN_eq_delta_psi}
    \cfrac{1}{r^2} \cfrac{d}{d r} \left( r^2 \cfrac{d \delta \psi_l}{d r} \right) -\cfrac{l(l+1)}{r^2} \delta \psi_l  
    = -4\pi G\rho^*\left(-\delta U_l+\delta\Pi_l \right) \\
    -4\pi G \left( 3\delta p_l - 3p \cfrac{ \delta \rho^*_l}{\rho^*} \right) -4\pi G\delta\rho^*_l\left(-U+\Pi+\cfrac{3p}{\rho^*} \right) \ ,
\end{multline}
and
\begin{equation} \label{pN_eq_deltaX}
    \cfrac{1}{r^2} \cfrac{d}{dr} \left( r^2 \cfrac{d \delta X_l}{d r} \right) - \cfrac{l(l+1)}{r^2} \delta X_l - 2 \delta U_l = 0 \ .
\end{equation}
In summary, we have six perturbation equations \eqref{pN_eq_perturbed_Euler_radial}--\eqref{pN_eq_deltaX} for six unknown variables $\xi^r_l, \delta p_l, \delta U_l, \delta V^r_l, \delta \psi_l$ and $ \delta X_l$. This allows us to formulate an eigenvalue problem for the oscillation mode frequency once we add the relevant boundary conditions.

\subsection{The dimensionless formulation}

In order to simplify the numerical calculations it is  advantageous to express the equations in dimensionless form. 
Following the spirit of the common formulation of the Newtonian problem (see \cite{1971Dziembowski}), we introduce the following dimensionless variables: 
\begin{equation} 
    y_1 = \cfrac{\xi_l^r}{r} \ ,
\end{equation}
\begin{equation}
      \ y_2 = \cfrac{1}{gr}  \cfrac{\delta p_l}{\rho} - \left[  1 + \cfrac{1}{c^2} \left( \Pi + \cfrac{p}{\rho^*} \right) \right] \cfrac{\delta U_l}{gr} \ , 
\end{equation}
the definition of which serves to decouple the $y_1$ and $y_2$ pair from $y_3$ and $y_4$ given by
\begin{equation} \label{definition_y3_y4}
     y_3 = \cfrac{\delta U_l}{g r} \ ,
\end{equation}
\begin{equation}
  y_4 = \cfrac{1}{g}\cfrac{{d}\delta U_l}{{d}r} \ .
 \end{equation}
 We also introduce
\begin{equation} 
    y_5 = \cfrac{\delta V_l^r}{gr^2} \ , 
\end{equation}
\begin{equation}
     y_6 = \cfrac{1}{gr}\cfrac{{d} \delta V_l^r}{{d}r} \ ,
\end{equation}
\begin{equation} 
    y_7 = \cfrac{\delta X_l}{gr^3} \ , 
\end{equation}
\begin{equation}
    y_8 = \cfrac{1}{g r^2}\cfrac{{d} \delta X_l}{{d}r} \ , 
\end{equation}
\begin{equation}
    y_9 = \cfrac{\delta \psi_l}{g^2 r^2} \ ,  
\end{equation}
and
\begin{equation}
    y_{10} = \cfrac{1}{g^2 r}\cfrac{{d} \delta\psi_l}{{d}r} \ .
\end{equation}
In these relations $g(r)$ is the local Newtonian gravitational acceleration defined by
\begin{equation} \label{define_g}
    g = \cfrac{GM_B(r)}{r^2} \ . 
\end{equation}
It is worth highlighting that this scaling impacts on the radial dependence of the variables. For example, given that the potential
$U$ has the same dimensions as $gr$, the motivation for defining $y_3$ by dividing by $gr$ instead of by $U$ is that, near the centre of the star $gr \rightarrow \mathcal{O}(r^2)$, which helps eliminate the divergence at the centre.
Another reason for working with this specific  scaling\footnote{There are, of course, different ways one may scale the variables. 
For example, one could replace $g(r)$ with a factor
\begin{equation}
    g_0 = \cfrac{4\pi}{3}G \rho_0 r \ ,
\end{equation}
where $\rho_0$ is the central mass density. This leads to a different set of dimensionless equations to solve, but the expansion near the centre remains the same. Ultimately, the strategy is a matter of choice.} is that this choice facilitates direct comparison with the corresponding Newtonian problem (e.g. as described in \cite{1979Unno}).

With the definitions of the dimensionless variables and \eqref{define_g}, the six perturbation equations above \eqref{pN_eq_perturbed_Euler_radial}-\eqref{pN_eq_deltaX} are reduced to ten first order differential equations for ten unknown variables. We have
\begin{multline} \label{ODEy1}
    r \cfrac{d y_1}{d r} = - \left( A^*   +  \cfrac{r d \rho^*}{\rho^* d r} + 3 \right) y_1   
    - V_g y_2   + \cfrac{l(l+1)}{ c_1 \tilde\omega^2 } \left[1 - \cfrac{1}{c^2} \left( \Pi + 4U  + \cfrac{p}{\rho^* }  \right) \right] y_2  \\
    - V_g \left[  1 + \cfrac{1}{c^2} \left( \Pi + \cfrac{p}{\rho^*} \right) \right] y_3  + \cfrac{gr}{c^2}  y_3  +  \cfrac{4gr}{c^2} \left(2  y_5 +  y_6 \right) - \cfrac{l(l+1) gr}{2 c^2} y_7 - \cfrac{l(l+1)}{c^2}  \cfrac{gr}{ c_1 \tilde\omega^2 } y_9  \ , 
\end{multline} 
\begin{multline} \label{ODEy2}
    r \cfrac{d y_2}{dr} = c_1 \tilde\omega^2 \left[1 + \cfrac{1}{c^2} \left( \Pi + 4U  + \cfrac{p}{\rho^*}  \right) \right]  y_1  + \cfrac{A^*}{\rho g} \cfrac{d p}{d r} y_1  -   \left( U_b - 1 - A^* \right)  y_2 +   \cfrac{1}{c^2 \rho^*}  \cfrac{d p}{d r}   r  y_2   \\
    - A^* \left[  1 + \cfrac{1}{c^2} \left( \Pi + \cfrac{p}{\rho^*} \right) \right] y_3  +  \cfrac{1}{c^2} \left( \cfrac{4}{\rho}\cfrac{d p}{d r} -  4\cfrac{d U}{d r} - \cfrac{d\Pi}{dr} +  \cfrac{p}{\rho^{*2}} \cfrac{d\rho^*}{dr} \right) r y_3 \\
    - c_1 \tilde\omega^2 \cfrac{4 gr}{c^2} y_5 +  c_1 \tilde\omega^2   \cfrac{gr}{2 c^2}  y_8   + \cfrac{gr}{c^2} y_{10} \ ,
\end{multline}
\begin{equation}
    r \cfrac{d y_3}{d r} =  \left( 1 - U_b \right) y_3 + y_4 \ ,
\end{equation}
\begin{equation}
    r \cfrac{d y_4}{d r} = - 4 \pi G \cfrac{\rho^* r}{g} \left( A^* y_1 + V_g y_2 \right) 
    - 4 \pi G \cfrac{\rho^* r}{g} V_g \left[  1 + \cfrac{1}{c^2} \left( \Pi + \cfrac{p}{\rho^*}  \right) \right] y_3  - 4\pi G \cfrac{3 \rho^* r^2}{c^2} y_3
    + l(l+1) y_3   - U_b y_4 \ ,
\end{equation}
\begin{equation}
     r \cfrac{d y_5}{d r} = -  U_b y_5 +  y_6  \ , 
\end{equation}
\begin{equation}
    r \cfrac{d y_6}{dr} = - \left( 3 + U_b \right) y_6 - \left[ 2-l(l+1) \right]  y_5  -  4\pi G \cfrac{\rho^* r}{g} y_1 -2 y_3 \ ,
\end{equation}
\begin{equation}
     r \cfrac{d y_7}{d r} = - \left( 1 + U_b   \right) y_7 +  y_8 \ ,
\end{equation}
\begin{equation}
    r \cfrac{d y_8}{d r} =  2 y_3 + l(l+1) y_7 -  \left( 2 + U_b   \right) y_8  \ ,
\end{equation} 
\begin{equation}
    r \cfrac{ d y_9}{d r} =  - 2 (U_b - 1) y_9  + y_{10} \ ,
\end{equation}
and
\begin{multline} \label{ODEy10}
     r \cfrac{d y_{10}}{d r} = 
     - 4\pi G \cfrac{\rho^* A^* }{g^2} \left( \cfrac{p}{\rho}   -  U + \Pi   \right) y_1 -  4\pi G  \left[ \cfrac{\rho^* r}{g}\cfrac{1}{\Gamma_1}  +  \cfrac{3 \rho r}{g} - \cfrac{\rho^*}{g^2} \left( U - \Pi \right) V_g \right] y_2 \\ 
     + 4\pi G   \cfrac{ r}{g} \left[  \rho^* +  \cfrac{3\rho}{c^2} \left( U - \Pi \right) \right] y_3 -  4\pi G   \left[ \cfrac{\rho^* r}{g}\cfrac{1}{\Gamma_1}  +  \cfrac{3 \rho r}{g} - \cfrac{\rho^*}{g^2} \left( U - \Pi \right) V_g \right] \left[  1 + \cfrac{1}{c^2} \left( \Pi + \cfrac{p}{\rho^*} \right) \right]  y_3 \\ + l(l+1) y_9 + \left( 1 - 2 U_b \right) y_{10}   \ .
\end{multline}
 In these equations we have used the additional definitions (again, similar to the Newtonian problem from \cite{1979Unno}):
\begin{equation}
    A^* = - rA =  - \cfrac{ r d \rho}{\rho d r} + \cfrac{r}{p\Gamma_1}\cfrac{d  p}{d r}  \ ,
\end{equation}
\begin{equation}
    V_g = \cfrac{\rho g r}{\Gamma_1 p} \ ,
\end{equation}
\begin{equation}
    U_b = \cfrac{d \ln M_{B}(r)}{d \ln r} = \cfrac{4\pi\rho^* r^3}{M_{B}(r)} = \cfrac{r}{g}\cfrac{d g}{d r} + 2 \ ,
\end{equation}
\begin{equation}
    c_1 = \left( \cfrac{r}{R} \right)^3 \cfrac{M}{M_{B}(r)} \ ,
\end{equation}
and we also introduce the scaled (dimensionless) frequency as
\begin{equation} \label{tildeomega}
    \tilde{\omega}^2 = \cfrac{\omega^2 R^3}{G M} \ .
\end{equation}
With these variables, the dimensionless formulation of the equations is complete.

\subsection{The boundary conditions}

Complemented by the appropriate boundary conditions at the centre and the surface of the star, the oscillation equations form an eigenvalue problem. The conditions we need to impose are natural extensions of the usual ones: First, we need the physical solution to be regular at the centre of the star. Second, at the star's surface we need the Lagrangian perturbation of the pressure to vanish while the potentials $\delta U_l$, $\delta X_l$, $\delta V_l^r$ and $\delta \psi_l$ and their derivatives are continuous across $r=R$. 

The analysis of the solution at the centre of the star is carried through a Taylor expansion. In order to avoid numerical difficulties, the integration of the equations is always initiated a small distance $r=r_{\epsilon}$ away from the origin.   
Adapting the strategy used  in \cite{1979Unno}, 
we  express the problem as a matrix problem 
\begin{equation} \label{ODEs}
    r \cfrac{d y_k}{d r} = \bar{A}_{kl} y_l
\end{equation}
where the matrix $\bar{A}_{lk}$ (provided in Appendix~\ref{AppB}) represents the central values  (actually, at $r_{\epsilon}$) of the various coefficients from the (dimensionless) perturbation equations. Then solving the characteristic equation
\begin{equation}
    \det (\bar{A}_{ij} - \lambda \delta_{ij} ) = 0 \ , 
\end{equation}
where the $\delta_{ij}$ is the Kronecker delta, we arrive at a set of  eigenvalues $\lambda_i$ and the corresponding eigenvectors $\bm{Y}_i$. Rejecting the singular solutions and combining the remaining ones one finds that (again, see Appendix~\ref{AppB} for details)   the following five conditions  must be satisfied  at the centre of the star:
\begin{equation}
    y_2 - \cfrac{1+l}{a} y_1 = 0 \ ,
\end{equation}
\begin{equation}
    y_4 - ly_3 = 0
\end{equation}
\begin{equation}
    y_6 - ( l -1 ) y_5 + \cfrac{3 y_1 + 2 y_3}{3+2l} = 0\ ,
\end{equation}
\begin{equation}
    y_8 - l y_7 - \cfrac{2 y_3}{3+2l} = 0 \ ,
\end{equation}
and
\begin{equation}
    y_{10} - l y_9 -  \cfrac{ae + (l+1)f}{a(3+2l)} y_1 - \cfrac{\bar{g}}{3+2l} y_3 = 0 \ . 
\end{equation}
Here we have defined (with subscript 0 indicating the central values of the various quantities)
\begin{equation}
    a  =  \cfrac{l(l+1)}{\omega^2}  \cfrac{4\pi}{3} G  \rho^*_0 \left[1 - \cfrac{1}{c^2} \left( \Pi_0 + 4U_0  + \cfrac{p_0}{\rho^*_0 }  \right) \right]  \ , 
\end{equation}
\begin{equation}
    c =  \cfrac{3 \omega^2}{4\pi G \rho^*_0} \left[1 + \cfrac{1}{c^2} \left( \Pi_0 + 4U_0  + \cfrac{p_0}{\rho^*0 }  \right) \right] \ ,
\end{equation}
\begin{equation}
    e = \left(  U_0 - \Pi_0 -\cfrac{p_0}{\rho_0}  \right)  \cfrac{9}{ 2\pi G  \rho^*_0    } \left( \cfrac{p_2}{p_0 \Gamma_{1}} - \cfrac{\rho_2}{\rho_0} \right)  \ ,
\end{equation}
\begin{equation}
    f = \left(  U_0 - \Pi_0  \right)  \cfrac{3}{\Gamma_{1}}\cfrac{\rho_0}{p_0}   -  \cfrac{3}{\Gamma_{1}} - 9 \left( 1 -\cfrac{3 U_0}{c^2} \right) \ ,
\end{equation}
and
\begin{equation}
    \bar{g} = 3 \left[ 1  +  \cfrac{3}{c^2} \left( U_0 - \Pi_0 \right)  \right] -  3 \left[   \cfrac{1}{\Gamma_1}  +  \cfrac{3 \rho_0}{\rho^*_0} - (U_0 - \Pi_0) \cfrac{\rho_0}{p_0\Gamma_1}  \right] \left[  1 + \cfrac{1}{c^2} \left( \Pi_0 + \cfrac{p_0}{\rho_0^*} \right) \right] \ .
\end{equation}
For realistic matter models, we should Taylor expand $\Gamma_1$ as well, but here we will only consider the simple case where $\Gamma_1$ is a constant.

Moving on to the behaviour at the stellar surface,   we need another five  conditions to be satisfied.  We  get the first of these conditions from $\Delta p = 0$. That is, we have
\begin{equation} \label{pN_bc_Delta_p_0}
    \Delta p_l = \delta p_l + \xi_l^r \cfrac{d p}{d r} = 0 \ .
\end{equation}
In the case of  the PW scheme the pressure gradient follows from \eqref{PWpressure} and (similar to the definition of $g$) we define
\begin{equation}
    g_1 = \cfrac{G\mathcal{N}}{r^2} \ ,  %\  \text{with} \ , % \ \mathcal{N} (r) = \int_0^r 4\pi \rho^* r'^2 \left(\Pi-U + \cfrac{3p}{\rho^*} \right)   dr' \ , 
\end{equation}
 to arrive at
\begin{equation}
    y_2 + \left[  1 + \cfrac{1}{c^2} \left( \Pi + \cfrac{p}{\rho^*} \right) \right] y_3 =  y_1 \left\{ \left[ 1 +   \cfrac{1}{c^2} \left( \Pi - 3U + \cfrac{p}{\rho^* }  \right)\right]  +  \cfrac{1}{c^2} \cfrac{g_1}{g} \right\} \cfrac{\rho^*}{\rho} \ .
\end{equation}
Given that $\Pi, \rho$ and $p$ vanish at the surface (for the stellar models we consider), we obtain a boundary condition constraining $y_1$ and $y_2$:
\begin{equation}
    y_2 = -  y_3 +  y_1 \left(  1 -   \cfrac{3U}{c^2}   +  \cfrac{1}{c^2} \cfrac{g_{1}}{g} \right) \left( 1 + \cfrac{3U}{c^2} \right) \    \text{at} \ r = R \ .
\end{equation}
 For the AGYM model, the equation for hydrostatic equilibrium is different and the boundary condition for $y_1$ and $y_2$ changes slightly. In this case we have:
\begin{equation}
    y_2  =  - y_3 + y_1 \left( 1 + \cfrac{g_{1}}{g} \right) \    \text{at} \ r = R  \ .
\end{equation}

In both cases, we also need to ensure the continuity of the four perturbed potentials $\delta U, \delta {\bm{V}}, \delta X$ and $\delta \psi$ and their derivatives across the surface. As discussed in Appendix~\ref{AppB}, this leads to the surface relations
\begin{equation}
    y_4 + (l+1) y_3 = 0 \ , 
\end{equation}
\begin{equation}
    (l+2)y_5 + y_6 - \cfrac{2 y_3}{2l-1} = 0 \ ,
\end{equation}
\begin{equation}
    (l+1) y_7 + y_8 + \cfrac{2}{2l-1} y_3 = 0 \ ,
\end{equation}
and
\begin{equation}
    y_{10} + (l+1) y_9 = 0 \ . 
\end{equation}
We now have all the conditions we need to impose in order to determine the required mode solutions. We may proceed to discuss the numerical results.

\section{Numerical results} \label{Sec_results}

The ten first order differential equations and ten boundary conditions together form an eigenvalue problem with the oscillation frequencies  the eigenvalues we want to determine. In order to solve the problem numerically, we adapt the strategy from the relativistic problem, see for instance  \cite{1983Lindblom, 1992Kokkotas}. Schematically, this involves integrating five linearly independent solutions that satisfy the conditions at the centre of the star (see Appendix~\ref{AppB}) and matching them to a set of five solutions obtained by integration backwards from the stellar surface (where they satisfy the required conditions (again, discussed in Appendix~\ref{AppB}).
Formally, representing the solution to the system of equations \eqref{ODEy1}-\eqref{ODEy10} by $\mathcal{Y} (r) = \left[ y_1, y_2, ..., y_{10}\right]^T$,  we have
\begin{equation} \label{mathcalY}
    r \cfrac{d}{dr} \mathcal{Y} = \bar{A} \mathcal{Y} \ .
\end{equation}
The boundary conditions at the centre provide five linearly independent vectors $\mathcal{Y}_i^c$ (with $i=1-5$) and the linear combination
\begin{equation}
    \mathcal{Y}_c = \sum_{i=1} ^5 a_i \mathcal{Y}_i^c  \ ,
\end{equation}
then provides the general solution. The coefficients $a_i$ are yet to be determined. Similarly, the solution that satisfies the required conditions at the surface can be expressed in terms of 
 $\mathcal{Y}_i^s$ (with $i=6-10$) each of which  satisfies the boundary conditions at the surface. The corresponding solution then takes the form 
\begin{equation}
    \mathcal{Y}_s = \sum_{i=6}^{10} a_i \mathcal{Y}_i^s  \ .
\end{equation}
Finally, the constant coefficients are determined by matching the two solutions at a suitable point inside the star. Choosing to match at the  middle of the star, $r_m=R/2$, we simply require  
\begin{equation} \label{matching}
    \mathcal{Y}_c (R/2) = \mathcal{Y}_s (R/2)\ .
\end{equation}
As discussed in, for example, \cite{2015Kruger.PhRvD..92f3009K} the matching condition can be turned into the requirement that the determinant constructed from  the numerical solution vectors must vanish. This then leads to the condition for the eigenvalues
\begin{equation}
    \det \left[ \mathcal{Y}_1^{c}, \mathcal{Y}_2^{c}, \mathcal{Y}_3^{c}, \mathcal{Y}_4^{c}, \mathcal{Y}_5^{c}, - \mathcal{Y}_6^{s}, - \mathcal{Y}_7^{s}, - \mathcal{Y}_8^{s}, - \mathcal{Y}_9^{s}, - \mathcal{Y}_{10}^{s} \right] = \mathcal{D}(\tilde{\omega}) = 0 \ .
\end{equation}
The values of $\tilde\omega= \tilde \omega_n$ for which the determinant vanishes are the mode frequencies. For each $\tilde\omega_n$ we can work out the eigenfunctions by solving \eqref{matching} for $a_1 - a_{10}$.

\begin{figure}
    \centering
    \includegraphics[width=0.49\textwidth]{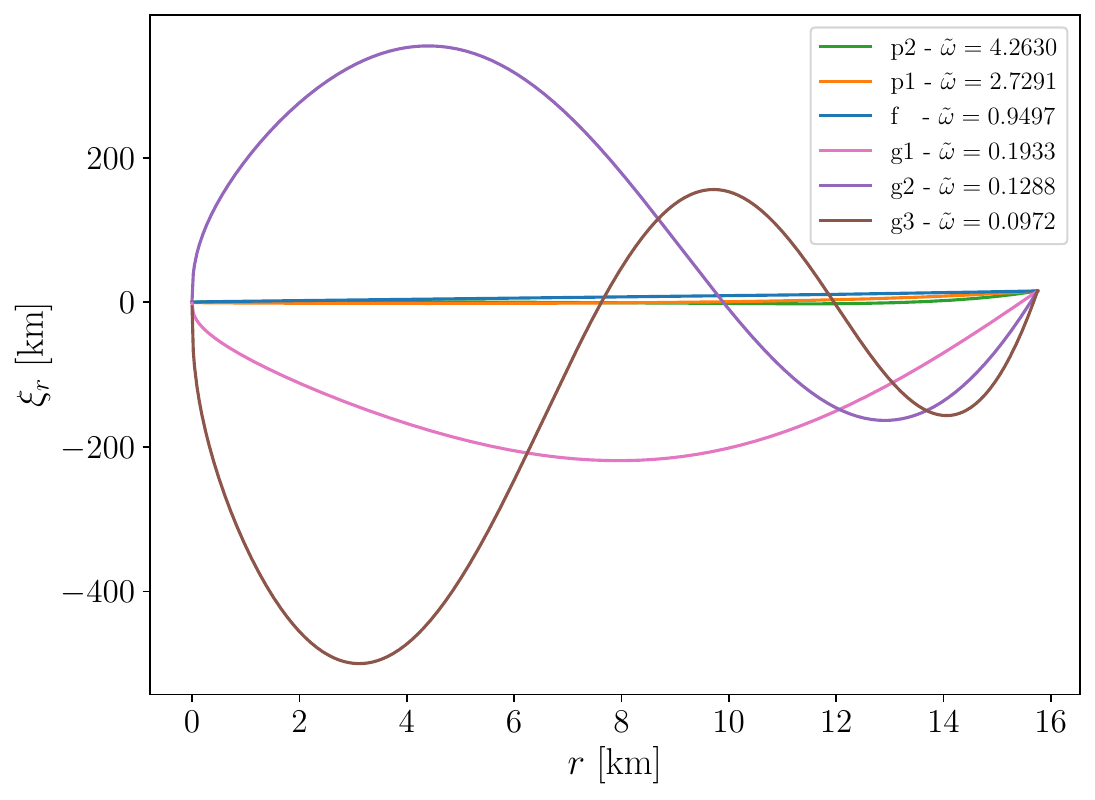} \includegraphics[width=0.49\textwidth]{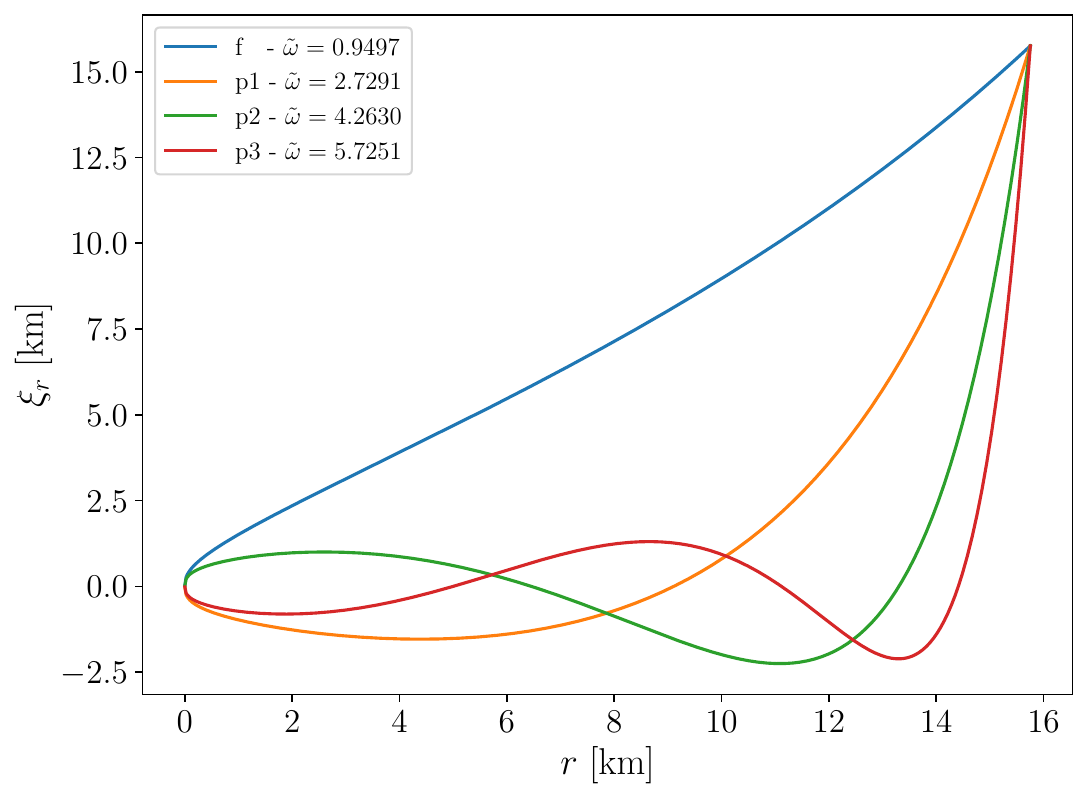} 
    \caption{The radial component of the displacement vector $\xi_r$ of $f, p_1, p_2, g_1, g_2, g_3$ modes for the static star with gravitational mass of $0.771 M_{\odot}$, built with PW models. These solutions are indicative of the behaviour found for low-mass stars in both pN formulations. }
    \label{PW3modes}
\end{figure}

In order to explore how the pN scheme performs, let us first consider three models with the relatively low baryon masses $M_B=0.436 M_{\odot},  0.669 M_{\odot} $ and $0.823 M_{\odot}$.
A sample of numerical results is provided in Table~\ref{Table:pNmodes}. As the table shows, we have calculated the fundamental f-mode and the lowest order p-modes and g-modes for a $\Gamma=2$ polytropic background model with constant adiabatic index $\Gamma_1=2.1$ for the perturbations. The details of the model are, however, not that important. The key points we want to stress here relate to the comparison of the different formulations of the problem. Specifically, Table~\ref{Table:pNmodes} provides results obtained for both PW and the AGYM background models. The data clearly bring out the expectation that the two models lead to similar results for low-mass stars. After all, in the weak-field limit the post-Newtonian corrections should be small. The numerical results also show that the mode frequencies tend to decrease when the gravitational redshift is accounted for, an effect that increases for more compact stars. Adding to this, we know from the mass-radius curves in Figure~\ref{PWfig} that our two pN models differ substantially in the stellar radius for a given mass as the stars become heavier. This highlights an obvious problem with this kind of comparison. 
In addition, the scaling of the frequency with the radius is problematic. In the pN scheme, the calculation is carried out using isotropic coordinates, while it is more common to solve the relativistic problem in Schwarzschild coordinates. In the Newtonian limit, this makes no difference but for heavier stars we need to pay attention to this. At the end of the day, these caveats suggest that the actual values for the frequencies provided in Table~\ref{Table:pNmodes} are less important than the trends in the results. The most important lessons are i) that the calculation can be carried out and the mode results for low-mass stars are sensible (including the eigenfunctions, see the exemplar provided in Figure~\ref{PW3modes}), but ii) the calculation unfortunately breaks for more massive stars. 

\begin{table}[!htb]
  \centering
  \begin{tabular}{|c|c|cc|cc|cc|}
  \hline
  $ $ & \textbf{Newtonian} &{\textbf{PW}} &\textbf{AGYM} & {\textbf{PW}}&\textbf{AGYM}  & {\textbf{PW}} & \textbf{AGYM}   \\
  \hline
  $M_{\textrm{B}}/M_{\odot}$ &  & 0.4361 &  0.4363 &0.6689 & 0.6689 & 0.8231 & 0.8232    \\
  $M/M_{\odot}$ & &  0.4247 &  0.4259 &  0.6374 & 0.6419 & 0.7712  & 0.8016   \\
  $R_{\textrm{S}}$/km &  & 16.5407 & 16.2712 &  16.6919 &  15.9570  &16.9930 & 15.5085    \\
  \hline
  $p_3$ & 7.4590   & 6.6291 & 6.7912 & 6.1016 & 6.4214 & 5.7251    & 5.9672  \\
  $p_2$ & 5.5743 & 4.9539  & 5.0775 & 4.5522 & 4.7951 & 4.2630  & 4.4461   \\
  $p_1$ & 3.5785 & 3.1849  & 3.2694 & 2.9211 & 3.0887 & 2.7291 & 2.8676  \\
  $\bm{f}$ & 1.2277 & 1.1232  & 1.1673& 1.0267   & 1.1237 & 0.9497 & 1.0756 \\
  $g_1$ & 0.2566 & 0.2314& 0.2406  & 0.2100 & 0.2304  & 0.1933  & 0.2165    \\
  $g_2$ & 0.1770 & 0.1582  & 0.1646 & 0.1417& 0.1557 & 0.1288  & 0.1439   \\
  \hline
  \end{tabular}

  \caption{The  dimensionless quadrupole ($l=2$) oscillation frequencies $\tilde\omega_n$ for our baseline polytropic model with $\Gamma=2, \ K=185$~km$^2$ and $\Gamma_1 = 2.1$. The numerical data bring out the expectation that the two pN models (PW and AGYM) agree well in the weak-field regime but, as the stellar radius corresponding to the same (baryon) mass begin to differ for more massive stars, the mode results also begin to diverge. In general, the mode frequencies decrease compared to the Newtonian case because of the gravitational redshift.}
\label{Table:pNmodes}
\end{table}

\begin{figure}
    \centering
    \includegraphics[width=0.7\textwidth]{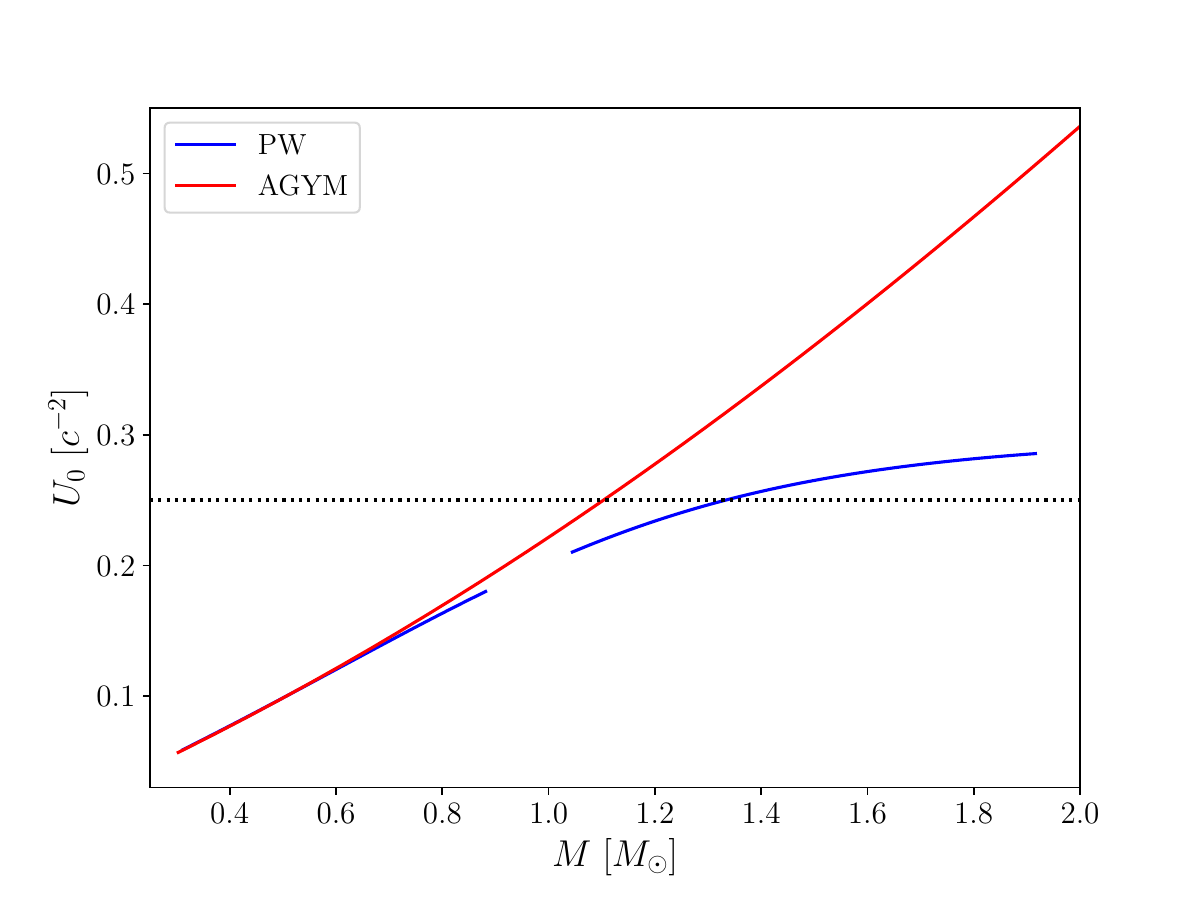} 
    \caption{The variation of the central value of the potential $U_0$ with the gravitational mass $M$ for stars built within the two pN schemes \eqref{PWpressure} (blue solid) and \eqref{Alter2pressure} (red solid) for the same polytropic equation of state as in Table~\ref{Table:pNmodes}.}
    \label{Uc_Mass}
\end{figure}

The problems associated with the pN scheme become apparent when we consider heavier (and hence more realistic) neutron star models. We know already from Figure~\ref{PWfig} that the  divergence of the PW scheme away from the fully relativistic background models is much more drastic than for the AGYM models, but the simple fact is that the error in the stellar radius is substantial for both schemes once we consider stars above $M_B\approx 1M_\odot$. This is not surprising as these stars are in the strong-gravity regime and one would intuitively expect the pN approach to break down. This is, indeed, what happens. However, it is interesting to note how and why the calculation breaks. We get an immediate clue to the answer from the results for a model with $M_B=1.322 M_{\odot}$. The eigenvalue calculation appears to proceed without a hitch, but when we consider the mode eigenfunctions we note that they are no longer well behaved near the origin. Specifically, the radial component of the displacement vector $\xi_r$ diverges at the centre. This behaviour can be understood from the discussion following equation~\eqref{save_centre} in Appendix~\ref{AppB}. If the central value of the gravitational potential, $U_0$,  becomes too large then the Taylor expansion no longer provides a well-behaved solution.  Evidence that this, indeed, happens is provided in Figure~\ref{Uc_Mass}. Once the value of $U_0$ exceeds 1/4 once would expect the calculation to be come problematic and this is exactly what happens. Again, it might be tempting to argue that this was entirely expected and that we do not learn very much from this exercise. Undoubtedly, this is true, but we suggest that it is nevertheless useful to make this result quantitative. It is now much clearer how the pN approach does not just become less accurate, it breaks completely, for realistic neutron star parameters. 

Having identified the problem---or at least the most pressing one---we can also ask if we can adjust the calculation to do better. We will suggest a strategy that would be helpful in this respect, but before doing so we should acknowledge that this involves abandoning that logic of the pN expansion. This inevitably involves a different set of choices and we are not going to claim anything other than that our adapted strategy provides a pragmatic way to circumvent the numerical problem we have identified. Schematically, noting that 
the divergence in the eigenfunctions appears near the origin, we consider what happens if we retain higher order pN terms in the small $r$ expansion.  After all, the results in Figure~\ref{Uc_Mass} clearly show that we cannot safely neglect the higher order terms involving the gravitational potential. 
One way to deal with this problem, inspired by discussion in \cite{1997Shibata, 2023Andersson}, is outlined in Appendix~\ref{IC_2PN}. 

The argument proceeds as follows: In the steps going from  \eqref{pN_eq_perturbed_Euler_radial}
to \eqref{ODEy2}, we multiplied by 
\begin{displaymath}
  \cfrac{1}{\rho g}\left( 1- \cfrac{2U}{c^2} \right) \ , 
\end{displaymath} 
expanded the result and kept only the 1pN terms. For example, we had
\begin{displaymath}
\left( 1- \cfrac{2U}{c^2} \right) \times \left( 1+ \cfrac{2U}{c^2} \right) A \cfrac{d p}{d r} r y_1 = \left( 1 -  \cfrac{4U^2}{c^4} \right) A \cfrac{d p}{d r} r y_1 \approx  A \cfrac{d p}{d r} r y_1 \ .
\end{displaymath} 
This simplification is evidently not valid in the  neutron star regime. 
A simple alternative would be, rather than expanding, to simply divide by the term including the gravitational potential. This leads to equations \eqref{ODEy2new} and \eqref{ODEy1new} for $y_2$ and $y_1$, respectively.
Replacing these two equations, we find that we are able extend the mode calculation to heavier masses without encountering any problems at the origin. Typical results obtained with this approach are provided in Table~\ref{Table:pNmodes2}. 
The models presented in the Table are indicated on the respective mass-radius curves in Figure~\ref{PWfig}.

\begin{table}[!htb]
  \centering
  \begin{tabular}{|c|c|cc|cc|cc|cc|cc|}
  \hline
  $ $ & \textbf{Newtonian} & \textbf{PW} & \textbf{AGYM} &  \textbf{PW} & \textbf{AGYM} & \textbf{PW} & \textbf{AGYM} & \textbf{PW} & \textbf{AGYM} & \textbf{PW} & \textbf{AGYM}  \\
  \hline
  $M_{\textrm{B}}/M_{\odot}$ &  & 0.4361 & 0.4363 & 0.6892 & 0.6892 & 0.8231 & 0.8232 & 1.3223 & 1.3224 & 1.5469 & 1.5469   \\
  $M /M_{\odot}$ &  & 0.4247 & 0.4259 & 0.6554 & 0.6604 & 0.7712 & 0.7801 & 1.1666 & 1.1961 & 1.3319 & 1.3674  \\
  $R_{\textrm{S}}$/km &  & 16.5407 & 16.2712 & 16.7222 & 15.9328 & 16.9930 & 15.7854 & 19.0227 & 15.4093 & 20.3351 & 15.3157   \\
  \hline
  $p_3$ & 7.4590 & 6.6881 & 6.8521 & 6.2115 & 6.5669 & 5.9567 & 6.4347 & 5.0484 & 6.0446 & 4.6819 &  5.9145  \\
  $p_2$ & 5.5743 & 5.0036 & 5.1288 & 4.6481 & 4.9193 & 4.4574 & 4.8222 & 3.7755 & 4.5369 & 3.4998 & 4.4424   \\
  $p_1$ & 3.5785 & 3.2209 & 3.3064 & 2.9916 & 3.1787 & 2.8667 & 3.1198 & 2.4109 & 2.9484 & 2.2237 & 2.8930    \\
  $\bm{f}$ & 1.2277 & 1.1376 & 1.1821 & 1.0522 & 1.1597 & 0.0968 & 1.1491 & 0.7376 & 1.1824  & 0.5984 & 1.1083   \\
  $g_1$ & 0.2566 & 0.2342 & 0.2434 & 0.2146 & 0.2373 & 0.2023 & 0.2345 & 0.1512 & 0.2268 & 0.1282 & 0.2245    \\
  $g_2$ & 0.1770 & 0.1611 & 0.1677 & 0.1472 & 0.1635 & 0.1385 & 0.1617 & 0.1024 & 0.1568 & 0.0867 & 0.1555    \\
  $g_3$ & 0.1360 & 0.1238 & 0.1289 & 0.1130 & 0.1257 & 0.1062 & 0.1243 &  0.0779 & 0.1206 & 0.0659 & 0.1197  \\
  \hline
  \end{tabular}
  \caption{The dimensionless mode oscillation frequencies $\tilde{\omega}_n$ for the reformulated pN problem. The stellar models are the same as in Table~\ref{Table:pNmodes}.}
  \label{Table:pNmodes2}
\end{table}

The new set of results  extend the calculation into the neutron-star mass range. Comparing to the data in Table \ref{Table:pNmodes}, we learn that the reformulation of the equations (evidently including higher order pN terms in a different way) lead to slightly higher frequencies for all modes. This effect becomes more pronounced for more massive stars, as one might expect given that the higher order pN terms play a more important role. Comparing the results for the PW and AGYM models we also see that---as expected, given the divergence of the two mass-radius curves, see Figure~\ref{PWfig}---the mode results for the PW model change sharply once the stellar mass is increased above $1M_\odot$.

\begin{figure}
    \centering
    \includegraphics[width=0.7\textwidth]{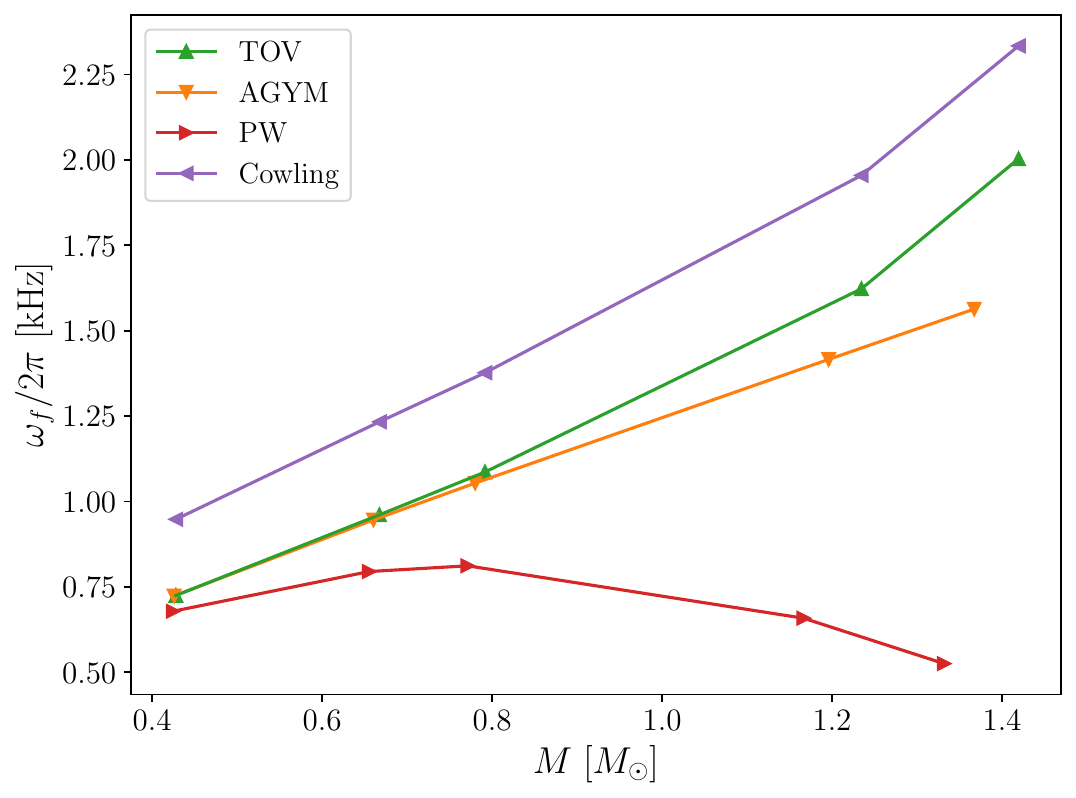}
    \caption{The f-mode frequencies $\omega_f/2\pi$ from Table~\ref{Table:GRmodes} shown as functions of the gravitational mass $M$ of the star. Here we also include results obtained within the PW formulation of the pN problem.}
    \label{fmode_mass}
\end{figure}

So far, we have mainly compared the two pN models to the Newtonian results. Given that the mode problem was formulated in a way that resembles the Newtonian problem (see, in particular, the version in \cite{1979Unno}) it made sense to make this comparison. Of course, the real test of the pN calculation must be a comparison to the fully relativistic problem. Having established that the calculation can be carried out for realistic neutron star masses, let us turn to this comparison. 
For this exercise we will no longer consider the dimensionless frequency $\tilde \omega$. Instead, as we do not want to bias the result by scaling with the radius (which we anyway know differs from the solution to the relativistic structure equations) we will consider the 
actual dimensional frequency $\omega/2\pi$ in kHz and a dimensionless representation $\bar \omega = G M \omega/c^3$ based on a scaling with the gravitational mass. 
We will also compare to results obtained within the relativistic Cowling approximation (where the perturbations of the spacetime metric, and hence the gravitational-wave aspects, are ignored). The corresponding results are obtained from the numerical code described in \cite{2024Rhys.MNRAS.531.1721C}. Moreover, we will focus our attention on the results  for the fundamental f-mode within the AGYM prescription. This is, after all, the ``best performing'' out of the pN frameworks we have considered. The relevant results are provided in Table~\ref{Table:GRmodes}. The corresponding PW results are included in Figure~\ref{fmode_mass}. The picture that emerges is this: The pN calculation provides very accurate results for low-mass stars. It is notably more precise than the relativistic Cowling calculation in this regime. The latter overestimates the mode frequencies by 15-20\% across the entire mass range we have considered. For stellar masses above about $1M_\odot$ the pN results become less accurate. For the most massive star we consider here ($M_B\approx 1.54 M_\odot$) the error in the pN mode frequency is similar to that of the Cowling calculation. Although, while the latter overestimates the frequency, the pN calculation underestimates it. The overall behaviour is nicely represented by Figure~\ref{fmode_mass}.

\begin{table}[!htb]
  \centering
  \begin{tabular}{|c|cc|cc|cc|cc|cc|}
  \hline
  $ $ & \textbf{TOV} & \textbf{AGYM}  & \textbf{TOV}   & \textbf{AGYM} & \textbf{TOV}    &\textbf{AGYM} & \textbf{TOV}      & \textbf{AGYM} &  \textbf{TOV}  & \textbf{AGYM}  \\
  \hline
  $M_{\textrm{B}}/M_{\odot}$ & 0.4365 &   0.4363  & 0.6895 &  0.6892  & 0.8234 &   0.8234 &   1.3223   & 1.3224 &   1.5445   & 1.5469  \\
  $M/M_{\odot}$ & 0.4279 &  0.4259  & 0.6675 &   0.6604 & 0.7917 &   0.7801  & 1.2346   & 1.1961    &  1.4195   & 1.3674 \\
  $R_{\textrm{S}}$/km & 16.0855 &   16.2712 & 15.4505 &   15.9328 & 15.0812 &   15.7854 &   13.3438   & 15.4093 &   12.0922   & 15.3157\\
  \hline
$(\omega_{\bm{f}}/2\pi)$/kHz &  0.7248 &   0.7234 & 0.9609 &    0.9460 & 1.0861 &   1.0537     & 1.6226   & 1.4160 &   2.0033 &  1.5633 \\
  Cowling  & 0.9478     & - & 1.2332   & - & 1.3767   & -   & 1.9546 & -    & 2.3337  & -  \\
  \hline
   $\bar\omega_{\bm{f}}$ & 0.009599 &   0.009535 & 0.01985 &    0.01933 & 0.02661 &  0.02544     & 0.05241  & 0.07898   & 0.08800 & 0.06615 \\
    Cowling  & 0.01255    & - & 0.02548 &   - & 0.03373 &   - &  0.07468  & -  &    0.1025 &  -\\
  \hline
  \end{tabular}
  \caption{Comparing  pN results for the fundamental mode (obtained from the AGYM formulation of the problem) to the fully relativistic results and  results obtained within the relativistic Cowling approximation. The pN models and the relativistic stars have the same baryon mass and hence can be considered to represent the ``same star'' in different representations of gravity.  }
  \label{Table:GRmodes}
\end{table}

\section{Concluding remarks} \label{Sec_summary}

With the aim of (eventually) describing the dynamical tides in  binary neutron star systems using a post-Newtonian mode-sum approach, we have developed a post-Newtonian perturbation formalism for the required neutron star dynamics. The  results we have provided convey a simple message: the pN formalism becomes less robust as the star becomes more compact. While it would be fair to suggest that this was expected---after all, the pN approximation assumes that the matter involves  slow motion, low pressures and a weak gravitational field, assumptions not relevant for neutron stars---there are important lessons to learn from the precise way in which the calculation breaks down. As our calculations demonstrate, the problem involves a number of subtle issues that warrant detailed investigation. Moreover, it is possible to tweak the formulation to avoid some of the numerical issues that arise.

As a positive note, our results show (see for example Figure \ref{fmode_mass}) how accurate the pN model is for lighter stars. In fact, the results demonstrate that the 
model remains useful up to a neutron star mass of about $1M_\odot$. 
We have also demonstrated that the pN calculation is much more accurate than the relativistic Cowling approximation in this regime.  This is, however, no longer the case for canonical $1.4M_\odot$ neutron stars, for which the error in the pN mode frequency is comparable to that of the Cowling calculation. The main take-home message is that a pN model provides a very accurate description for the  dynamics of mildly relativistic systems (e.g. white dwarf oscillation modes, tides etcetera) and may provide useful insights into the dynamics of low-mass neutron stars, as well.

Having carried out this investigation, we  have a much better understanding of how the various weak-field assumptions associated with pN theory impact on the problem of  stellar oscillations. While we started with an intuitive idea of the problem, we now have a quantitative picture. Specifically, we know at what point the 
pN background model becomes dubious and how this breakdown influences the dynamics at the linear perturbation level. With a more precise knowledge of the limitations of---and conceptual challenges associated with---the pN approach, we may consider future applications, like the development of a mode-sum approach for dynamical tides, with open eyes.

\section*{Acknowledgments}

SY wish to thank Rhys Counsell and David Trestini for helpful discussions. NA gratefully acknowledges support from the STFC via grant No.~ST/Y00082X/1.
F.G. acknowledges funding from the European Union’s Horizon Europe research and innovation programme under the Marie Sk{\l}odowska-Curie grant agreement No.~101151301.

\appendix

\section{Thermodynamical relations} \label{thermal_perturbation}

In this Appendix we show that equation \eqref{deltaPi_internal_energy}, which is used to express the perturbed internal energy, holds for both barotropic and non-barotropic (frozen composition) models. 

Let us first consider the barotropic case. For a single fluid, with $\varepsilon=\varepsilon(n)$, the usual Gibbs relation leads to
\begin{equation}
    p + \varepsilon = n \cfrac{d\varepsilon}{dn} \ ,
    \label{thermrel}
\end{equation}
where $n$ is the baryon number density. Introducing the (baryon) mass density, $\rho = mn$ (with $m$ the rest mass of each baryon), we therefore have
\begin{equation} \label{gibbs}
    \cfrac{d\varepsilon}{d\rho} = \cfrac{p+\varepsilon}{\rho} \ .
\end{equation}
In the pN problem we introduced the internal energy per unit mass $\Pi$ through \eqref{introduce_internal_energy}. This means that  \eqref{gibbs} leads to
\begin{equation} 
    d\Pi - \cfrac{p}{\rho^2 } d\rho = 0 \ .
    \label{1stlaw}
\end{equation}
The same relation holds for Eulerian perturbations, so we have the required result: 
\begin{equation}
    \delta\Pi = \cfrac{p}{\rho^2 } \delta\rho \ .
\end{equation}

The non-barotropic case is a little bit more involved. Assuming that the matter is cold enough that we can ignore thermal effects, we start from  a two-parameter equation of state, say $\varepsilon=\varepsilon(n,x_\text{p})$, where the second parameter is the proton fraction $x_\text{p}$. In general, we need to account for nuclear reactions \cite{2019Andersson.Pnigouras.MNRAS.489.4043A,2024MNRAS.531.1721C}. For an npe (neutrons, protons and electrons) system we have
\begin{equation} \label{Gibbs_npe}
p+\varepsilon = n_\text{n} \mu_\text{n} + n_\text{p} \mu_\text{p} + n_\text{e} \mu_\text{e} \ ,
\end{equation}
where each chemical potential $\mu_\mathrm{x}$ is defined by
\begin{equation}
    \mu_\text{x} =  \cfrac{\partial \varepsilon}{\partial n_\text{x}}  \ , \quad \mathrm{x=n,p,e}\ .
\end{equation}
For the equilibrium  configuration, we impose beta-equilibrium which means that
\begin{equation}
    \mu_\text{n} = \mu_\text{p} + \mu_\text{e} \ . 
\end{equation}
We also require local charge neutrality, so 
\begin{equation}
    n_\text{p} = n_\text{e} \ .
\end{equation}
Imposing the latter condition and introducing $n=n_\mathrm{n}+n_\mathrm{p}$ we have 
\begin{equation} \label{Gibbs2}
    p+ \varepsilon  = n\mu_\text{n} + n x_\text{p} (\mu_\text{p}+\mu_\text{e}-\mu_\text{n}) \ ,
\end{equation}
which, in equilibrium, leads back to \eqref{thermrel} once we identify $\mu = \mu_\mathrm{n}(n)$.
Similarly, a variation of the energy gives (in terms of $\rho$ rather than $n$)
\begin{equation}
  d\varepsilon = {1\over m} \left[ (1-x_\text{p})\mu_\text{n} + x_\text{p}(\mu_\text{p} + \mu_\text{e}) \right] d\rho +  {\rho \over m} (\mu_\text{p}  + \mu_\text{e} - \mu_\text{n} ) dx_\text{p} \ .
\end{equation}

Meanwhile, from the form of the energy assumed in the pN calculation \eqref{introduce_internal_energy} we should have
\begin{equation}
    d\varepsilon =   (c^2 + \Pi)  d\rho+ \rho \left( {\partial \Pi \over \partial \rho}  \right)_{x_\text{p}} d\rho +  \rho \left( {\partial \Pi \over \partial x_\text{p}}  \right)_{\rho} dx_\text{p} \ .
\end{equation}
We are interested in frozen composition---i.e., assume that nuclear reactions are too slow to equilibrate that matter on the timescale associated with the dynamics---which means that the variations ensure that $dx_\text{p}=0$. Assuming that the background model is in chemical equilibrium, we then arrive at the relation 
\begin{equation} \label{partialPi_partialrho}
    \left( {\partial \Pi \over \partial \rho}  \right)_{x_\text{p}}  = {p\over \rho^2} \ .
\end{equation}

In order to tie everything together, consider
 Lagrangian perturbations for which  we have
\begin{equation}
    \Delta p = \left( {\partial p \over \partial \rho}\right)_{x_\text{p}} \Delta \rho + \left( {\partial p \over \partial x_\text{p}}\right)_{\rho} \Delta x_\text{p} \ .
\end{equation}
Provided the nuclear reactions are slow enough, we assume the composition to be frozen, so $\Delta x_\text{p}=0$ and 
\begin{equation}
       \Delta p = \left( {\partial p \over \partial \rho}\right)_{x_\text{p}} \Delta \rho \equiv {p\Gamma_1 \over \rho} \Delta \rho \ . 
\end{equation}
This provides the thermodynamical definition for the adiabatic index $\Gamma_1$. For the Eulerian perturbations, it follows that
\begin{equation}
    \delta p ={p\Gamma_1 \over \rho} \delta \rho + \left( {p\Gamma_1 \over \rho} \xi^i \partial_i \rho - \xi^i \partial_i p\right) \ ,
\end{equation}
In terms of the internal energy, with frozen composition, we must have
\begin{equation}
    \Delta \Pi =  \left( {\partial \Pi\over \partial \rho}\right)_{x_\text{p}} \Delta \rho \ ,
\end{equation}
so
\begin{equation} 
  \delta \Pi =\left( {\partial \Pi\over \partial \rho}\right)_{x_\text{p}} \delta \rho + \left[\left( {\partial \Pi\over \partial \rho}\right)_{x_\text{p}} \xi^i \partial_i \rho - \xi^i \partial_i \Pi\right] = \left( {\partial \Pi\over \partial \rho}\right)_{x_\text{p}} \delta \rho + \left[\left( {\partial \Pi\over \partial \rho}\right)_{x_\text{p}} \ - {p\over \rho^2}  \right] \xi^i \partial_i \rho \ ,
\end{equation}
which is consistent with the barotropic result.
Finally, making use of \eqref{partialPi_partialrho} we arrive at
\begin{equation}
    \delta \Pi = \cfrac{p}{\rho^2} \delta \rho \ ,
\end{equation}
which is equation \eqref{deltaPi_internal_energy} from the main body of the paper.

%-------------
\section{Boundary conditions}
\label{AppB}

\subsection{At the centre of the star}

In order to work out the required boundary conditions at the centre of the star---essentially, imposing regularity of the perturbations---we make use of a Taylor expansion. For the background quantities, this means that we have (at an initial point $r=r_{\epsilon}$ near the centre)
\begin{equation*}
    p = p_0 + p_2 r^2 + ... \ , \ \rho = \rho_0 + \rho_2 r^2 +... \ .
\end{equation*}
We also know that, for the polytropic model we are working with we have
\begin{equation}
    p_0 = K \rho_0^2 \quad \mbox{and}\quad \ p_2 = 2 K \rho_2 p_0 \ .
\end{equation}
It is straightforward to work out the corresponding behaviour of the centre for all other background quantities. For example, one finds that 
\begin{equation}
    p_2 = - G \cfrac{2\pi}{3}\rho^{*2}_0 \left[  1 + {2 \over c^2} \left( \Pi_0 - 2U_0 + { 3 p_0 \over \rho^*_0 } \right)   \right] \ ,
\end{equation} 
with
\begin{equation}
  \rho^*_0 = \rho_0 \left( 1 + \cfrac{3U_0}{c^2} \right) \ .  
\end{equation}

The analysis of the perturbation equations is more involved. Expressing the set of equations as a matrix problem
\begin{equation} 
    r \cfrac{d y_k}{d r} = \bar{A}_{kl} y_l \ ,
\end{equation}
the coefficient matrix $\bar{A}_{lk}$ is given by 
\begin{equation}
    \bar{A}_{ij} = 
    \begin{bmatrix}
-3 & a & 0 & 0 & 0 & 0 & 0 & 0 & 0 & 0 \\
c & -2 & 0 & 0 & 0 & 0 & 0 & 0 & 0 & 0 \\
0 & 0 & -2 & 1 & 0 & 0 & 0 & 0 & 0 & 0 \\
0 & 0 & l(l+1) & -3 & 0 & 0 & 0 & 0 & 0 & 0 \\
0 & 0 & 0 & 0 & -3 & 1 & 0 & 0 & 0 & 0 \\
-3 & 0 & -2 & 0 & -2+l(l+1) & -6 & 0 & 0 & 0 & 0 \\
0 & 0 & 0 & 0 & 0 & 0 & -4 & 1 & 0 & 0 \\
0 & 0 & 2 & 0 & 0 & 0 & l(l+1) & -5 & 0 & 0 \\
0 & 0 & 0 & 0 & 0 & 0 & 0 & 0 & -4 & 1 \\
e & f & \bar{g} & 0 & 0 & 0 & 0 & 0 & l(l+1) & -5 
\end{bmatrix}
\label{Amat}
\end{equation}
with
\begin{equation}
    a  =  \cfrac{l(l+1)}{\omega^2}  \cfrac{4\pi}{3} G  \rho^*_0 \left[1 - \cfrac{1}{c^2} \left( \Pi_0 + 4U_0  + \cfrac{p_0}{\rho^*_0 }  \right) \right]  \ , 
\end{equation}
\begin{equation}
    c =  \cfrac{3 \omega^2}{4\pi G \rho^*_0} \left[1 + \cfrac{1}{c^2} \left( \Pi_0 + 4U_0  + \cfrac{p_0}{\rho^*_0 }  \right) \right] \ ,
\end{equation}
\begin{equation}
    e = \left(  U_0 - \Pi_0 -\cfrac{p_0}{\rho_0}  \right)  \cfrac{9}{ 2\pi G  \rho^*_0    } \left( \cfrac{p_2}{p_0 \Gamma_{1}} - \cfrac{\rho_2}{\rho_0} \right)  \ ,
\end{equation}
\begin{equation}
    f = \left(  U_0 - \Pi_0  \right)  \cfrac{3}{\Gamma_{1}}\cfrac{\rho_0}{p_0}   -  \cfrac{3}{\Gamma_{1}} - 9 \left( 1 -\cfrac{3 U_0}{c^2} \right) \ ,
\end{equation}
and
\begin{equation}
    \bar{g} = 3 \left[ 1  +  \cfrac{3}{c^2} \left( U_0 - \Pi_0 \right)  \right] -  3 \left[   \cfrac{1}{\Gamma_1}  +  \cfrac{3 \rho_0}{\rho^*_0} - (U_0 - \Pi_0) \cfrac{\rho_0}{p_0\Gamma_1}  \right] \left[  1 + \cfrac{1}{c^2} \left( \Pi_0 + \cfrac{p_0}{\rho_0^*} \right) \right] \ ,
\end{equation}
Note that, for a realistic matter model we should Taylor expand $\Gamma_1$ as well, but we only consider the simpler case with $\Gamma_1$ constant.

At this point we have a choice to make. Assuming a power law solution $y_i=c_i r^\lambda$, we can use computer algebra to solve the matrix problem
\begin{equation}
    \det (\bar{A}_{ij} - \lambda \delta_{ij} ) = 0 \ , 
\end{equation}
where  $\delta_{ij}$ is the Kronecker delta, to obtain the set of  eigenvalues $\lambda_i$ and the corresponding eigenvectors $\bm{Y}_i$.

This way we arrive at the 10 eigenvalues
\begin{multline}
    \lambda_1 =  \cfrac{1}{2} \left(-5 - \sqrt{1+4ac} \right) \ , \ \lambda_2 = \cfrac{1}{2} \left(-5 + \sqrt{1+4ac} \right) \ , \ \lambda_3 = -5 - l \ , \ \lambda_4 = -5 - l \ , \  \lambda_5 =  -5 - l \ ,  \\
    \lambda_6 = -3 - l \ , \ \lambda_7 = - 4 + l \ , \ \lambda_8 = -4 + l \ , \ \lambda_9 = -4 + l \ , \ \lambda_{10} = - 2 + l \ .
\end{multline}
Moreover, we have 
\begin{equation} \label{ac_2pN}
    ac = l(l+1) \left[1 - \cfrac{1}{c^4} \left( \Pi_0 + 4U_0  + \cfrac{p_0}{\rho^*_0 }  \right)^2 \right]\approx  l(l+1) \ ,
\end{equation}
so it follows that
\begin{equation}
    \lambda_1 = -3 - l \ , \ \lambda_2 = -2 + l \ .
\end{equation}
At this point it would seem as if the eigenvalue problem is degenerate--- given that we have repeated eigenvalues---so  we have to proceed with care. This issue is discussed in \cite{1979Unno} but we find their analysis somewhat misleading. 
The coefficient matrix from \eqref{Amat} clearly hints at a block-diagonal structure and it is, in fact, straightforward to  demonstrate that the equations can be decoupled to show that  (even though the eigenvalues are degenerate) the 
corresponding eigenvectors are linearly dependent. 

Discarding the 5 singular solutions (for the physical variables!), we can express the  solution to \eqref{ODEs} as 
\begin{equation}
    y_i = \sum_{k=1}^{5} c_k \bm{Y}_k^i r^{\lambda_i} \ , 
\end{equation}
where $c_k$ are constants and $\bm{Y}_k^i$ represents the $i-$component of the eigenvector associated with the eigenvalue $\lambda_k$. Slightly abusing the notation, we relabel the eigenvalues in such a way that $\lambda_2 \to \lambda_1$, $\lambda_7 \to \lambda_2$, $\lambda_8\to\lambda_3$, $\lambda_9\to \lambda_4$ and $\lambda_{10}\to \lambda_5$ and identify the associated  the associated eigenvectors 
\begin{equation}
    \bm{Y}_1 =
    \begin{bmatrix}
        \cfrac{  (3 + 2l) 2a  }{(2+l)(ae +  fl+ f)} \\
        \cfrac{2(3 + 5 l + 2 l^2)}{ (2+l) ( ae + fl+ f )} \\
        0 \\
        0 \\
        \cfrac{ - 3a }{ (2 + l)(ae  +  fl + f) }  \\
        \cfrac{- 3 a  (1+l)}{(2+l)(ae + fl + f)}  \\
        0 \\
        0 \\
        \cfrac{2}{3+\sqrt{1+4ac}} = \cfrac{1}{2+l}\\
        1
    \end{bmatrix}  \ ; \
    \bm{Y}_2 =
    \begin{bmatrix}
    0 \\
    0 \\
    0 \\
    0 \\
    0 \\
    0 \\
    0 \\
    0 \\
    1/l \\
    1
    \end{bmatrix}  \ ;  \ 
    \bm{Y}_3 =
    \begin{bmatrix}
    0 \\
    0 \\
    0 \\
    0 \\
    0 \\
    0 \\
    1/l \\
    1 \\
    0 \\
    0
    \end{bmatrix}  \ ; \
    \bm{Y}_4 =
    \begin{bmatrix}
    0 \\
    0 \\
    0 \\
    0 \\
    \cfrac{1}{l-1} \\
    1 \\
    0 \\
    0 \\
    0 \\
    0
    \end{bmatrix}  \ ; \ 
    \bm{Y}_{5} =
    \begin{bmatrix}
    0 \\
    0 \\
    \cfrac{2(3 + 2l )}{ \bar{g}(2+l)  }  \\
    \cfrac{2(3 + 2l )l}{ \bar{g}(2+l) } \\
    \cfrac{ -2 }{ \bar{g}(2+l)  } \\
    \cfrac{ -2 (1 + l )}{ \bar{g}(2+l)  } \\
    \cfrac{ 2}{ \bar{g}(2+l)  } \\
    \cfrac{ 2}{ \bar{g}} \\
    \cfrac{1}{2+l} \\
    1 \\
    \end{bmatrix}  \ .
\end{equation}

Expressing the solution in terms of this basis one can show that the following relations need to be imposed at the centre of the star:
\begin{equation}
    y_2 - \cfrac{1+l}{a} y_1 = 0 \ ,
\end{equation}
\begin{equation}
    y_4 - ly_3 = 0 \ ,
\end{equation}
\begin{equation}
    y_6 - ( l -1 ) y_5 + \cfrac{3 y_1 + 2 y_3}{3+2l} = 0 \ ,
\end{equation}
\begin{equation}
    y_8 - l y_7 - \cfrac{2 y_3}{3+2l} = 0 \ ,
\end{equation}
\begin{equation}
    y_{10} - l y_9 -  \cfrac{ae + fl + f}{a(3+2l)} y_1 - \cfrac{\bar{g}}{3+2l} y_3 = 0 \ . 
\end{equation}

In principle, this completes the argument. However, in order to fully understand the solutions it is worth taking a closer look a how the equations can be decoupled. In order to illustrate the strategy, let us focus on the first of the five solutions. 

Consider, first of all, the equations for $y_1$ and $y_2$ (obtained from the $2\times 2$ block in the upper left corner of \eqref{Amat}).
The eigenvalues associated with these equations require
\begin{equation}
    \left| \begin{array}{cc} \lambda+3 & - a \\  - c & \lambda+2  \end{array} \right| = 0 \ ,
\end{equation}
or
\begin{equation}
    \left( \lambda +3  \right) \left( \lambda + 2 \right) -  ac  = 0 \ ,
    \label{quadra}
\end{equation}
where $ac$ is given by \eqref{ac_2pN}. Discarding the higher order pN terms, we see that the two roots are $l-2$ and $-l-3$. The first gives the regular solution at the centre and the corresponding eigenvector is such that
\begin{equation}
    y_1  = \cfrac{a}{l+1} y_2 =  \cfrac{l}{\omega^2} \cfrac{4\pi G}{3} \rho^*_0   \left[1 - \cfrac{1}{c^2} \left(  \Pi_0 +4 U_0  + \cfrac{p_0}{\rho^*_0}  \right) \right]      y_2 \ .
\end{equation}
 
 With this relation in hand, we note from \eqref{Amat} that we need particular integrals for some of the other variables. Clearly, the solution is consistent with $y_3=y_4=0$, but the $y_5$ and $y_6$ solutions have to be related in such a way that 
\begin{equation}
(l+1) y_5 =  y_6  
\end{equation}
and
\begin{equation}
   \left[ (l+4)(l+1) - l(l+1) + 2 \right] y_5 = 3 y_1 = \cfrac{3a}{l+1}    y_2 \ ,
\end{equation}
These relations lead to
\begin{equation}
    y_5 = \cfrac{3}{2(2l+3)}\cfrac{a}{l+1} y_2  \ . 
\end{equation}

Finally, there is no coupling to $y_7$ or $y_8$, but $y_9$ and $y_{10}$ are linked by 
\begin{equation}
    (l+2) y_9 = y_{10} \ ,
\end{equation}
and
\begin{equation}
    \left[ (l+2) (l+3)   - l(l+1) \right]  y_9 = 2(2l+3) y_9 = e y_1 + f y_2 = \cfrac{ae}{l+1} y_2 + f y_2 \ . 
\end{equation}
This completes one of the independent solutions to the system. It is, in fact, easy to show that the argument reproduce (up to a constant factor) the  $\bm{Y}_2$ solution from before.

The step-by-step argument is useful because it allows us to make an additional point. 
Let us go back to \eqref{quadra} but this time without neglecting the higher order pN contributions. We then have 
to solve the quadratic
\begin{equation} \label{save_centre}
    \left( \lambda +3  \right) \left( \lambda + 2 \right) -   l(l+1) \left[1 - \cfrac{1}{c^4} \left( \Pi_0 + 4U_0  + \cfrac{p_0}{\rho^*_0 }  \right)^2 \right]  = 0 \ .
\end{equation}
The roots now become
\begin{equation}
    \lambda = -{5\over 2} \pm {1\over 2} \left\{1 + 4l(l+1) \left[1 - \cfrac{1}{c^4} \left( \Pi_0 + 4U_0  + \cfrac{p_0}{\rho^*_0 }  \right)^2 \right]  \right\}^{1/2} \ .
\end{equation}
This becomes problematic if the argument of the square root changes sign. A necessary condition for this to happen is
\begin{equation}
    \cfrac{1}{c^4} \left( \Pi_0 + 4U_0  + \cfrac{p_0}{\rho^*_0 }  \right)^2 > 1 \ .
\end{equation}
Ignoring the contributions from the pressure and the internal energy, this corresponds to
\begin{equation}
{4U_0\over c^2} \gtrsim 1 \ . 
\end{equation}
This clearly never happens in the weak-field regime, but as we demonstrate in the main text the central value of the gravitational potential  becomes large enough for this to be an issue when we consider realistic neutron star parameters.

\subsection{At the surface of the star}

As mentioned in the main text, at the star's surface we need the Lagrangian perturbation of the pressure to vanish while the potentials $\delta U_l$, $\delta X_l$ and $\delta \psi_l$ and their derivatives are continuous across $r=R$. Here show more details of the boundary condition for $\delta U_l$ and $\delta V_l$. Starting from \eqref{pN_eq_delta_U}, 
\begin{equation*} 
    \cfrac{1}{r^2} \cfrac{d}{d r} \left( r^2 \cfrac{d \delta U_l}{d r} \right) -\cfrac{l(l+1)}{r^2} \delta U_l + 4\pi G \rho^* \left( \cfrac{\delta p_l}{\Gamma_1 p} -  \xi_l^r A   + \cfrac{3 \rho}{c^2 \rho^*}\delta U_l \right) = 0 
\end{equation*}
outside the star $\rho^* = 0$ so we have
\begin{equation} \label{BC_deltaU_1}
    \cfrac{1}{r^2} \cfrac{d}{d r} \left( r^2 \cfrac{d \delta U_l}{d r} \right) -\cfrac{l(l+1)}{r^2} \delta U_l = 0 \ ,
\end{equation}
which leads to
\begin{equation}
 \delta U_l= A r^{-l-1}
\end{equation}
where $A$ is a constant. Then $\delta U_l$ satisfies the condition
\begin{equation}
    \cfrac{d \delta U_l}{d r} + \cfrac{l+1}{r} \delta U_l = 0 \quad \text{at} \ r = R \ ;
\end{equation}
which leads to 
\begin{equation}
    y_4 + (l+1) y_3 = 0 \ .
\end{equation}
Note that this is the same as in Newtonian case.

Next, for the vector potential $\delta V_l^r$, the equation with $\rho^*$ vanishing outside the star, turns into 
\begin{equation} \label{pN_eq_delta_hat_Ur(r)_outside}
    \cfrac{1}{r^2} \cfrac{d}{d r}\left(r^2 \cfrac{d \delta V^r_l}{d r}\right) + \cfrac{2}{r} \cfrac{d \delta V^r_l}{d r}  + \cfrac{2 - l(l+1)}{r^2}  \delta V^r_l = - \cfrac{2}{r} \delta U_l \ ,
\end{equation}
i.e., a non-homogeneous linear differential equation. The general solution should be the sum of the particular solution and the solution to the corresponding complementary equation. As $\delta U_l$ behaves like $r^{-l-1}$, we can then tell from \eqref{pN_eq_delta_hat_Ur(r)_outside} that the particular solution should have the  form
\begin{equation}
    \delta V^r_{l \ \text{p}} = B_1 r^{-l} \ .
\end{equation}
We can get a relation between the constant $A$ and $B_1$ by substituting the particular solution back into the equation:
\begin{equation}
    B_1 (2l-1) = A \ .
\end{equation}
Assuming the solution to complementary equation of \eqref{pN_eq_delta_hat_Ur(r)_outside}
\begin{equation}
    \delta V^r_{\text{c}} = B_2 r^{\beta} \ ,
\end{equation}
substituting into the complementary equation
\begin{equation*}
    \beta (\beta+1) r^{\beta} + 2\beta r^{\beta} + \left[ 2-l(l+1) r^\beta \right] = 0
\end{equation*}
we find that
\begin{equation}
    \beta = l+1 , \ -l-2
\end{equation}
where $\beta = -l-2$ ensures that $\delta V_l^r$ is regular at infinity for $l \geq 2$. Therefore the solution to $\eqref{pN_eq_delta_hat_Ur(r)_outside}$ should take the form
\begin{equation} \label{BC_for_delta_hat_Ur_1}
    \delta V_l^r = B_1 r^{-l} + B_2 r^{-l-2} = \cfrac{A}{(2l-1)r^l} + \cfrac{B_2}{r^{l+2}} \ , 
\end{equation}
while its derivative is
\begin{equation} \label{BC_for_delta_hat_Ur_2}
    \cfrac{d \delta V_l^r}{d r} = - B_1 l r^{-l-1} - (l+2) B_2 r^{-l-3} = - \cfrac{Al}{(2l-1)r^{l+1}} - \cfrac{B_2(l+2)}{r^{l+3}} \ .
\end{equation}
Eliminating the constant $B_2$ from \eqref{BC_for_delta_hat_Ur_1}\eqref{BC_for_delta_hat_Ur_2}, we arrive at
\begin{equation}
    (l+2)y_5 + y_6 = \cfrac{2 y_3}{2l-1}  \ .
\end{equation}
Similarly, we can work out the conditions for the potentials $\delta X_l$ and $\delta \psi_l$.

\section{Keeping higher order terms} \label{IC_2PN}

Having identified the issue associated with the Taylor expansion at the centre of the star, we want to see if there is an ``easy'' fix to the numerical problem. One possible strategy proceeds as as follows: In the steps going from  \eqref{pN_eq_perturbed_Euler_radial}
to \eqref{ODEy2}, we multiplied by 
$$\cfrac{1}{\rho g}\left( 1- \cfrac{2U}{c^2} \right)$$ expanded the result and kept only the 1pN terms. For example, we had
$$\left( 1- \cfrac{2U}{c^2} \right)\times \left( 1+ \cfrac{2U}{c^2} \right) A \cfrac{d p}{d r} r y_1 = \left( 1 -  \cfrac{4U^2}{c^4} \right) A \cfrac{d p}{d r} r y_1 \approx  A \cfrac{d p}{d r} r y_1 \ . $$ 
This last step is evidently not valid in the neutron star regime. 
A simple alternative is to, rather than expanding, simply divide by the term including the gravitational potential.  Then, in place of \eqref{ODEy2} we have for $y_2$:
\begin{multline} 
     r \cfrac{d y_2}{d r} = 
    c_1 \Tilde{\omega}^2  \left[1 + \cfrac{1}{c^2} \left( \Pi + 3 U  + \cfrac{p}{\rho^* }  \right) \right] \cfrac{\rho^*}{\rho} \left( 1 + \cfrac{2U}{c^2} \right)^{-1} y_1 +  \cfrac{A^*}{\rho g} \cfrac{d p}{d r} y_1 \\
    + \left( A^* - \cfrac{r}{g} \cfrac{d g}{d r} - 1 \right) y_2  + \cfrac{1}{\rho^* c^2}  \cfrac{d p}{d r} \left( 1 + \cfrac{2U}{c^2} \right)^{-1} r y_2  \\
    + A^* \left[  1 + \cfrac{1}{c^2} \left( \Pi + \cfrac{p}{\rho^*} \right) \right] y_3 - \cfrac{1}{c^2} \left( \cfrac{d \Pi}{d r}  + \cfrac{1}{\rho^*} \cfrac{d p}{d r} - \cfrac{p}{\rho^{*2}}\cfrac{d \rho^*}{d r} \right) r y_3 \\
    + \cfrac{1}{\rho^* c^2}  \cfrac{d p}{d r} \left( 1 + \cfrac{2U}{c^2} \right)^{-1} r \left[  1 + \cfrac{1}{c^2} \left( \Pi + \cfrac{p}{\rho^*} \right) \right] y_3 
    +  \cfrac{4}{\rho c^2} \left(\cfrac{d p}{d r} - {\rho^*} \cfrac{d U}{d r}  \right) \left( 1 + \cfrac{2U}{c^2} \right)^{-1}r y_3 \\
    + \cfrac{\rho^*}{\rho} \left[1 + \cfrac{1}{c^2} \left(  \Pi - U  + \cfrac{p}{\rho^* }  \right) \right] \left( 1 + \cfrac{2U}{c^2} \right)^{-1} y_4 - \left[1 + \cfrac{1}{c^2} \left(  \Pi   + \cfrac{p}{\rho^* }  \right) \right]  y_4       \\
    - \cfrac{\rho^*}{\rho} \left( 1 + \cfrac{2U}{c^2} \right)^{-1} \left( c_1 \Tilde{\omega}^2  \cfrac{4 gr}{c^2}  y_5 - c_1 \Tilde{\omega}^2    \cfrac{gr}{ 2 c^2}  y_8   - \cfrac{1}{c^2}  g r y_{10} \right) \ .
    \label{ODEy2new}
\end{multline}
Similarly for $y_1$, instead of \eqref{ODEy1}, we have:
\begin{multline}
    r \cfrac{d y_1}{dr} = - \left( A^*  +  \cfrac{r}{\rho^*}\cfrac{d \rho^*}{d r}  +  3   \right) y_1  
    - V_g  y_2  
    + \cfrac{\rho}{\rho^*} \left(1 + \cfrac{2U}{c^2} \right) \left[1 + \cfrac{1}{c^2} \left(  \Pi + 3 U  + \cfrac{p}{\rho^* }  \right) \right]^{-1}  \cfrac{l(l+1)}{c_1 \Tilde{\omega}^2 } y_2 \\
     - V_g \left[  1 + \cfrac{1}{c^2} \left( \Pi + \cfrac{p}{\rho^*} \right) \right] y_3 + \cfrac{\rho}{\rho^*} \left(1 + \cfrac{2U}{c^2} \right) \left[  1 + \cfrac{1}{c^2} \left( \Pi + \cfrac{p}{\rho^*} \right) \right] \left[1 + \cfrac{1}{c^2} \left(  \Pi + 3 U  + \cfrac{p}{\rho^* }  \right) \right]^{-1}  \cfrac{l(l+1)}{c_1 \Tilde{\omega}^2 }   y_3 \\
     + \cfrac{gr}{c^2} \left[1 + \cfrac{1}{c^2} \left(  \Pi + 3 U  + \cfrac{p}{\rho^* }  \right) \right]^{-1} y_3 - \left[1 + \cfrac{1}{c^2} \left(  \Pi -  U  + \cfrac{p}{\rho^* }  \right) \right]\left[1 + \cfrac{1}{c^2} \left(  \Pi + 3 U  + \cfrac{p}{\rho^* }  \right) \right]^{-1} \cfrac{l(l+1)}{c_1 \Tilde{\omega}^2 }  y_3 \\
      + \left[1 + \cfrac{1}{c^2} \left(  \Pi + 3 U  + \cfrac{p}{\rho^* }  \right) \right]^{-1} \left[    \cfrac{4 gr}{c^2} \left( 2 y_5 + y_6 \right) 
     - \cfrac{gr}{2 c^2} l(l+1) y_7 - \cfrac{gr}{c^2} \cfrac{l(l+1)}{c_1 \Tilde{\omega^2}} y_9  \right]    \ . 
     \label{ODEy1new}]
\end{multline}
Using these two equations in place of the original ones we find that the numerical problems at the centre of the star can be avoided.

\bibliography{Ref}

\end{document}